\documentclass[pra,aps,tightenlines,twocolumn,showpacs,amssymb,amsfonts,superscriptaddress]{revtex4}

\newcommand{\ket}{\rangle}

\usepackage{epsfig}

\begin{document}

\title{Bitwise Bell inequality violations for an entangled state involving $2N$ ions}
\author{D. T. Pope}
\email{d.pope@griffith.edu.au}
\affiliation{School of Physical
Sciences, Centre for Quantum Computer Technology, University of
Queensland, Brisbane 4072, Queensland, Australia}
\affiliation{Centre for Quantum Dynamics, School of Science,
Griffith University, Brisbane 4111, Australia}
\author{G. J. Milburn}
\email{milburn@physics.uq.edu.au} \affiliation{School of Physical
Sciences, Centre for Quantum Computer Technology, University of
Queensland, Brisbane 4072, Queensland, Australia}

\date{\today}

\begin{abstract}
Following on from previous work [J.-\AA. Larsson, Phys. Rev. A
{\bf 67}, 022108 (2003)], Bell inequalities based on correlations
between binary digits are considered for a particular entangled
state involving $2N$ trapped ions. These inequalities involve
applying displacement operations to half of the ions and then
measuring correlations between pairs of corresponding bits in the
binary representations of the number of centre-of-mass phonons of
$N$ particular ions. It is shown that the state violates the
inequalities and thus displays nonclassical correlations. It is
also demonstrated that it violates a Bell inequality when the
displacements are replaced by squeezing operations.
\end{abstract}


\maketitle

\section{Introduction} \label{intro}

%
%general background
%

Entangled quantum states typically, if not always, exhibit
nonclassical correlations. These correlations are crucial elements
in most quantum information processing tasks including quantum
computation \cite{nielsen00,preskill98}, quantum teleportation
\cite{bennett93}, superdense coding \cite{bennett92} and some
forms of quantum cryptography \cite{nielsen00} (Section 12.6).
Given this significance, it is important to consider how to best
observe such correlations and thus better understand the quantum
resources present in certain situations. One way of observing
nonclassical correlations is via the violation of Bell
inequalities \cite{bell64,clauser78}. For example, violations of
the Clauser-Holt-Shimony-Horne (CHSH) inequality \cite{clauser69}
can reveal the presence of such correlations in many two-qubit
entangled states. Similarly, a violation of the GHZ inequality
\cite{ghz89,mermin90} highlights the existence of nonclassical
correlations in the state $| \psi_{\rm GHZ} \rangle = 1/\sqrt{2}
\left( |000 \rangle  +  |111\rangle \right)$.
%In this instance,
%the inequality is violated by an appreciable amount, revealing the
%existence of highly nonclassical correlations.
Finally, the violation of a Bell inequality involving
higher-dimensional spin \cite{mermin80} by the spin-$s$ singlet
state, where $s=3/2, 5/2, 7/2 \ldots$ highlights nonclassical
correlations present in this state. Many other Bell violations are
also known, however, they are too numerous to mention.

%
%not all Bell inequalities are good for all states
%

The examples in the previous paragraph involved Bell inequalities
well-suited to observing nonclassical correlations in particular
entangled states. However, not all Bell inequalities are useful
for observing such correlations in every entangled state. For
instance, applying the CHSH inequality to any two qubits in
$|\psi_{\rm GHZ}\rangle$ produces no Bell violation and hence,
when used in this manner, this inequality does not highlight
$|\psi_{\rm GHZ}\rangle$'s nonclassical correlations. Similarly,
the W state \cite{dur00} $| \psi_{\rm W} \rangle = 1/\sqrt{3}
\left( |001 \rangle  + |010 \rangle + |01\rangle \right)$
satisifies the GHZ inequality and hence this inequality is
ill-suited for observing its nonclassical correlations. Another
noteworthy point about Bell violations and entangled states is
that certain mixed entangled states, namely bound entangled
states, may not violate {\em any} Bell inequality as has been
conjectured by Peres \cite{peres99}. Consistent with this, it has
been shown that multipartite bound entangled states for which all
partial transposes are positive satisfy one particular Bell
inequality \cite{werner00}.

%
%general key concept
%

Given that individual Bell inequalities can be either good or bad
tools for observing nonclassical correlations in specific
entangled states, it seems interesting to consider the following
question: ``Which particular Bell inequalities are best suited to
observing the nonclassical correlations of a certain entangled
state?'' Whilst not addressing this general question in the
current paper, we do show that certain Bell inequalities involving
correlations between binary digits in the binary representations
of particular observables can be used to detect interesting
nonclassical correlations in a particular entangled state
involving two sets of $N$ ions.
%
%explaining what Larrson did
%
In doing so, we follow on from \cite{larrson03} which showed that
the steady-state intracavity state of the nondegenerate parametric
amplifier (NOPA)
\begin{equation}
| {\rm NOPA} \rangle = \frac{1}{\cosh r} \sum_{n=0}^{\infty}
\tanh^{n} r | n \rangle_{1} | n \rangle_{2},
\end{equation}
where $|n\rangle$ is a photon number state, the subscripts 1 and 2
denote the signal and idler modes and $r$ is a real squeezing
parameter, violated certain Bell inequalities. More specifically,
using an {\em abstract mathematical} scheme \cite{larrson03}
showed that if we consider the numbers of photons in the signal
and idler modes in binary (eg. $|n=3\rangle_{1} \mapsto |n=\ldots
00011\ket_{1}$) then each pair of corresponding bits in the two
binary representations simultaneously violates a CHSH Bell
inequality. That is, it showed that the least significant bits for
the signal and idler modes, together, violated such an inequality
as did the second least significant bits, the third least
significant bits and so forth. The paper \cite{larrson03} also
briefly suggested how we might observe these violations but, on
this point, remarked that a better (that is, presumably, a more
experimentally achievable) measurement scheme than that suggested
was desirable \cite{larrson03} (p. 022108).

%
%Larsson's results are preceded by others
%

The results in \cite{larrson03} can be seen as extending those in
\cite{banasek98,banasek99a,banasek99b,chen02,gour03} which all
showed that $| {\rm NOPA} \rangle$ violated Bell inequalities
based on measuring photon-number parity (oddness or even-ness). In
terms of binary representations, these other papers violated Bell
inequalities involving the values of the least significant bits
(and not any other bits as did \cite{larrson03}) in the binary
representations of the numbers of photons in the signal and idler
modes.

%
%what's in this paper
%

This paper extends and complements work in \cite{larrson03} by
explicitly showing the existence of Bell violations closely
related to those in \cite{larrson03} within a {\em tangible} and
{\em arguably experimentally feasible} context that is different
to the context suggested in \cite{larrson03}. In particular,
motivated by \cite{larrson03}'s comment (p. 022108) that the
formulation of a practical measurement scheme for the Bell
inequalities in \cite{larrson03} is desirable, we (arguably)
propose such a scheme.
%we violate Bell inequalities based on \cite{larrson03}'s but,
%unlike \cite{larrson03} which mainly considered its Bell
%inequalities in an abstract manner, we consider ours in relation
%to a concrete measurement scheme.
This paper also extends work in \cite{larrson03} by illustrating
different ways to violate the sorts of Bell inequalities in
\cite{larrson03} to the ways shown in \cite{larrson03}. Instead of
violating Bell inequalities by measuring a range of pseudo-spin
observables for the state $| {\rm NOPA} \ket$ as did
\cite{larrson03}, we apply a range of displacements and squeezing
operations to a state generated from $| {\rm NOPA} \ket$ and then
always measure the same pseudo-spin observables.

%
%what's in this paper
%

The current paper proceeds as follows: in Section~\ref{state} the
state we consider is described, along with the physical system
underlying it which centres around two sets of $N$ trapped ions.
In Section~\ref{scheme}, the Bell inequalities we consider are
presented by outlining the measurements and operations they
involve. The measurements consist of measuring bits in the binary
representations of ${\cal N}_{1}$ and ${\cal N}_{2}$, where ${\cal
N}_{j}$ ($j=1,2$) is the number of centre-of-mass phonons for one
of the sets of $N$ ions in the $x$ direction, whilst the
operations (which are applied prior to the measurements) are
displacements applied to the centre-of-mass vibrational states of
one of the sets of ions. In Section~\ref{results}, it is shown
that the entangled state violates the Bell inequalities. Next,
Section~\ref{ch4:local_squeeze} presents a Bell inequality
involving local squeezing operations which the entangled state
also violates. Finally, the paper concludes with a discussion of
its results in Section~\ref{discussion}.

%
%why this paper is interesting
%

\section{The entangled state} \label{state}
\begin{figure}[h]
\center{\epsfig{figure=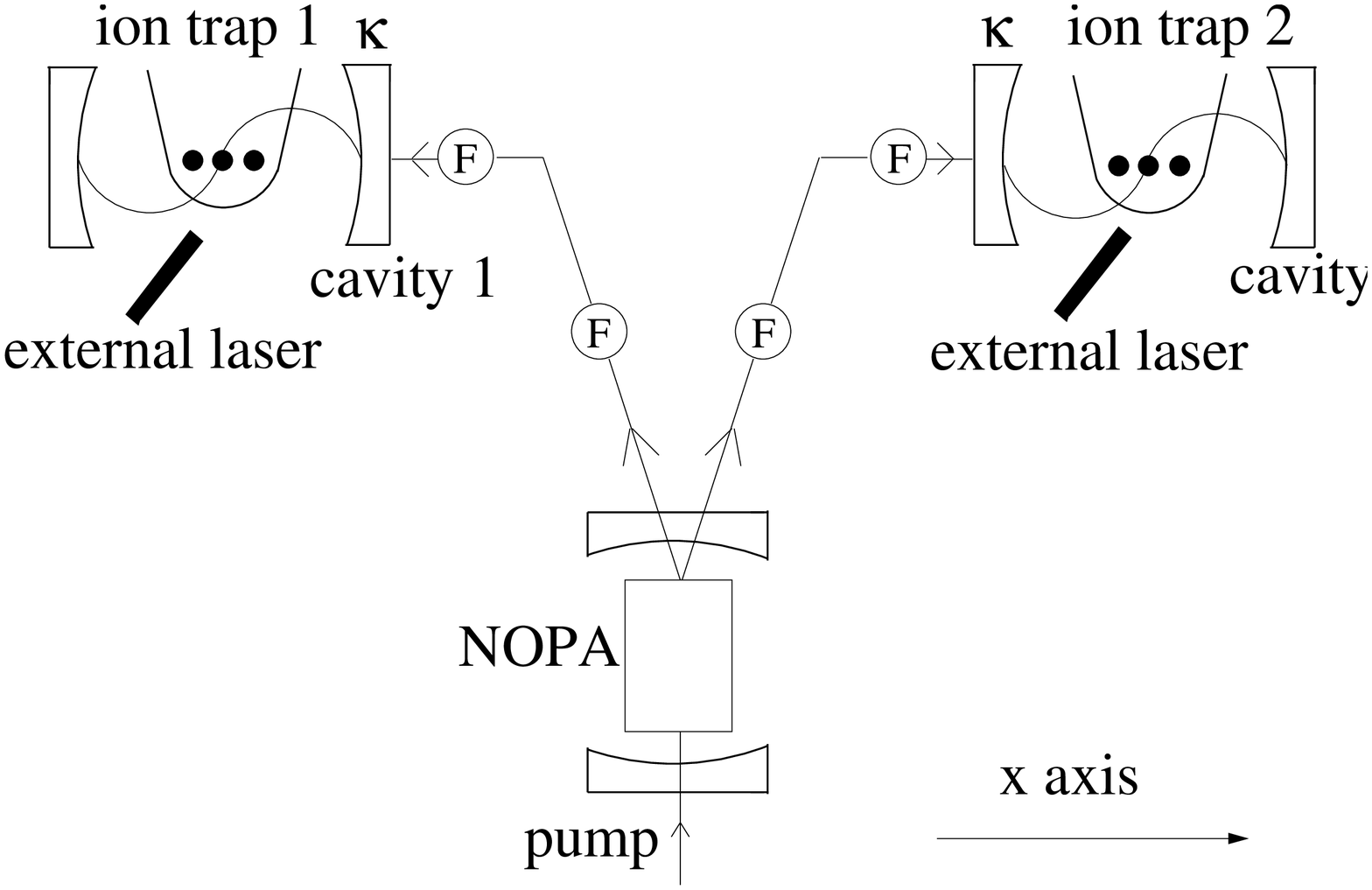,width=80mm}}
\caption{\label{ch3:fig_two} Schematic diagram for the system
associated with the $2N$-way entangled state. The system consists
of, firstly, a subthreshold optical nondegenerate parametric
amplifier (NOPA) whose output modes pass through Faraday isolators
(represented by an F enclosed in a circle) and then feed into
linearly damped optical cavities. These cavities are aligned along
the $x$-axis and each has one ideal mirror and one lossy one (with
damping constant $\kappa$). Inside each cavity is a harmonic ion
trap that confines $N$ identical two-level ions (each represented
by a black circle) in a linear chain parallel to the $x$-axis.
External lasers of frequency $\omega_{L}$ are incident on the
first ions in both traps from a direction perpendicular to the
$x$-axis.}
\end{figure}

The system associated with the entangled state considered
in this paper comprises of  a nondegenerate parametric amplifier
(NOPA) \cite{ou92a,ou92b,kimble92} and two linear ion traps which
each lie within an optical cavity and contain $N$ identical ions.
A schematic diagram of this system is shown in
Fig.~\ref{ch3:fig_two}. The NOPA operates below threshold and its
two external output fields first pass through Faraday isolators. Each of
them then feeds into a different linearly-damped optical cavity
via a lossy mirror. The cavities are aligned such that their axes
coincide with the $x$ axis and are closed at one end by perfectly
reflecting mirrors. In addition, each cavity supports a cavity
mode of frequency $\omega_{c}$ described by the annihilation
operator $a_{j}$, where $j$ enumerates the cavities. Within both
cavities lie $N$ identical two-level ions of mass $M$, charge $Z$
and internal transition frequency $\omega_{a}$. These ions are
trapped in a linear configuration parallel to the $x$-axis by a
harmonic potential (a linear ion trap \cite{ion_traps}) and hence
are tightly confined in the $y$ and $z$ directions. Furthermore,
the vibrational motion of the $m^{\rm th}$ ion in the $j^{\rm th}$
trap in the $x$ direction is described by the annihilation
operator $b_{jx}^{(m)}$ for which $[b_{jx}^{(m)},
b_{jx}^{(m)\;\dag}]=1$. The traps are aligned
such that the $j^{\rm th}$ trap is centred on a node of the cavity
field described by $a_{j}$. Finally, external lasers of frequency
$\omega_{L}$ whose beams are perpendicular to the $x$-axis are
incident on the first ions of both traps.

%
%Hamiltonian for a cavity
%

The Hamiltonian for the $j^{\rm th}$ optical cavity, the ions within
it and its reservoir is
\begin{eqnarray} \nonumber
H_{\rm j\:total} & = & H_{j0}^{\rm ion} + H_{j0} + H_{jI}^{\rm ion-ion} +
H_{jI} \\
& & + \kappa (a_{j} R^{\dag}_{j} + a^{\dag}_{j} R_{j}) + H_{\rm j\:res},
\end{eqnarray}
where $H_{j0}^{\rm ion}$ is the free Hamiltonian for
the vibrational states of the ions and
$H_{j0}$ is the free Hamiltonian for the cavity field and the ions' internal
states. The term $H^{\rm ion-ion}_{jI}$ describes the electromagnetic coupling between ions
whilst $H_{jI}$  describes a Raman process involving the
cavity field, the external laser and the first ion in the $j^{\rm th}$ trap.
Finally, $H_{\rm j\:res}$ is the Hamiltonian for the external reservoir
coupled to the $j^{\rm th}$ cavity for which $R_{j}$ is a reservoir annihilation operator and
$\kappa$ is a damping constant. More precisely,
$H_{j0}^{\rm ion} = \hbar \nu_{x} \sum_{m=1}^{N} \left(
b_{jx}^{(m)\:\dag} b_{jx}^{(m)} +\frac{1}{2}\right)$,
where $\nu_{x}$ ($\nu_{x}=\omega_{c} - \omega_{L}$) is the frequency
of both harmonic traps along the $x$-axis.
The Hamiltonian $H_{j0}$ is, in a frame rotating at frequency $\omega_{L}$,
\begin{equation} \label{free_hamiltonian}
H_{j0} = \hbar \delta a_{j}^{\dag} a_{j} + \hbar \Delta \sum_{m=1}^{N} \sigma_{j+}^{(m)}
\sigma_{j-}^{(m)},
\end{equation}
where $\delta = \omega_{c} -\omega_{L}$, $\Delta=\omega_{a} - \omega_{L}$,
and $\sigma_{j+}^{(m)}$ and $\sigma_{j-}^{(m)}$ are
raising and lower operators for the internal states of the $m^{\rm
th}$ ion in the $j^{\rm th}$ trap. The term $H_{jI}^{\rm ion-ion}$ is
\cite{james}
\begin{equation}
H_{jI}^{\rm ion-ion} = \Sigma_{m,n=1;\: m \neq n}^{N}
\frac{Z^{2}}{8 \pi \epsilon_{0} |x_{jn}(t)-x_{jm}(t) |},
\end{equation}
where $\epsilon_{0}$ is the permittivity of free space
and $x_{jl}$, for $l=1 \ldots N$, is the position of the $l^{\rm th}$ ion in the $j^{\rm th}$ trap.
The interaction term $H_{jI}$ is
\begin{eqnarray} \nonumber
H_{jI}& = & \hbar \big[{\cal E}_{L}(y,z,t) \sigma_{j+}^{(1)}+{\cal E}^{*}_{L}(y,z,t)
\sigma_{j-}^{(1)} \bigr] \\
& & + \hbar g_{0} \sin (k x_{j1}) \bigl( a^{\dag}_{j} \sigma_{j-}^{(1)}
+a_{j} \sigma_{j+}^{(1)} \bigr),
\end{eqnarray}
where ${\cal E}_{L}$ is the complex amplitude for
both external lasers, $k = \omega_{c} / c$ and $g_{0}$ (
$g_{0} \in \Re$) is the coupling constant for the ion-field
interaction.

%
%asssumptions about model
%

The following feasible assumptions are made about the system
\cite{parkins99} in order to simplify calculations and to focus on
its most important aspects:
\begin{enumerate}
\item All ions are so cold that they only move from their mean
position $x_{jl}^{0}$ by a small amount and so we can approximate
$x_{jl}(t)$ by $x_{jl}^{0}+q_{jl}(t)$, where $q_{jl}(t)$ is a
small displacement. \item The cavity field and external laser
frequencies are appreciably detuned from $\omega_{a}$ and all
two-level ions are initially in their ground states. Thus, the
excited internal states are sparsely populated and spontaneous
emission effects are negligible and can be ignored. \item The
wavelength of the cavity mode is much greater than the distances
that the first ions in both traps stray from the centres of their
traps and thus $\sin (kx_{j1}) \simeq k x_{j1}  <<  1$. This
allows us to to arrange things so that the $y$ and $z$ dependences
of the external laser fields are negligible and thus, assuming
${\cal E}_{L}$ is time independent, that ${\cal E}_{L}(y,z,t)
\simeq {\cal E} e^{-i\phi_{L}}$, where ${\cal E}$ is a real
time-independent amplitude.
\item The damping parameter $\kappa$
is such that $\nu_{x} \gg \kappa \gg g_{0} k {\cal E}
\sqrt{\hbar}/ \left( \sqrt{2 M \nu_{x} N} \Delta \right)$.
\item
For each trap, the frequencies of different normal or collective
modes \cite{james} in the $x$-direction are well-separated. Thus,
the cavity modes only couple to the centre-of-mass modes in this
direction.
\end{enumerate}

%
%simplify H
%

Given assumptions 1, 3 and 5, calculations in \cite{james}
show that we can write $H_{\rm j\:total}$ in terms of normal-mode
creation and annihilation operators as
\begin{eqnarray} \nonumber \label{com_eqn}
H_{\rm j\:total} & = & \hbar \sum_{m=1}^{N} \nu_{m} \left( B_{jx}^{\dag\:(m)} B_{jx}^{(m)}+\frac{1}{2}
\right) + H_{j0} \\ \nonumber
& & +
\hbar \big[{\cal E}_{L}(y,z,t) \sigma_{j+}^{(1)}+{\cal E}^{*}_{L}(y,z,t)
\sigma_{j-}^{(1)} \bigr] \\ \nonumber
& & + \frac{\hbar g_{0} \eta_{x}}{\sqrt{N}} (B_{jx}^{(1)}+B_{jx}^{(1)\:\dag})
\bigl( a^{\dag}_{j} \sigma_{j-}^{(1)} +a_{j} \sigma_{j+}^{(1)} \bigr) \\
& & + \kappa(a_{j} R^{\dag}_{j} + a^{\dag}_{j} R_{j}) + H_{\rm j\:res},
\end{eqnarray}
where $B_{jx}^{(m)}$ is the annihilation operator for the $m^{\rm
th}$ normal mode for the $j^{\rm th}$ trap in the $x$ direction.
For example, $B_{jx}^{(1)}$ is a centre-of-mass mode annihilation
operator which is $B_{jx}^{(1)} = 1/\sqrt{N} \left( b_{jx}^{(1)} +
b_{jx}^{(2)} + \ldots b_{jx}^{(N)} \right)$ whilst $B_{jx}^{(2)}$
is the annihilation operator for the {\em breathing mode} which is
$B_{jx}^{(2)} = 1/\sqrt{2} \left( -b_{jx}^{(1)} +
b_{jx}^{(2)}\right)$ when $N=2$. Observe that in
Eq.~(\ref{com_eqn}) the cavity mode only couples to the
centre-of-mass vibrational mode in the $x$ direction.

%the system is not overly impractical

Though it would be very challenging, at best, to experimentally
realise the system outlined above, it is potentially feasible to
do so. This is because, first, optical cavities and parametric
oscillators have been widely realised in laboratories. Second,
recent experiments \cite{mundt02} have trapped single ions in
electromagnetic traps lying within optical cavities.

%
%by analogy qusex occurs into COM mode
%

Upon adiabatically eliminating the cavity mode $a_{j}$ and also
$\sigma_{j+}$ and $\sigma_{j-}$ in Eq.~(\ref{com_eqn}), it can be
shown that the Langevin equation for $B_{jx}^{(1)}$ is
\begin{equation} \label{collective_langevin}
{\dot B}_{jx}^{(1)} = - \left(  \frac{\Gamma}{N} + i\nu_{x}\right) B_{xj}^{(1)} - \sqrt{\frac{2\Gamma}{N}}e^{-i \nu_{x}t} a_{\rm in},
\end{equation}
where ${\dot x}$ denotes the partial derivative of $x$ with
respect to time, $\Gamma = \hbar g_{0}^{2}  k^{2} {\cal E}^{2} /
\left( 2  M \nu_{x} \kappa \Delta^{2} \right)$ and $a_{\rm in}$ is
a quantum noise operator \cite{gardiner85}. This equation shows
that the only effect of having multiple ions, as opposed to a
single ion, in the trap is to introduce a factor of $1/N$ in front
of $\Gamma$. From \cite{parkins99} (especially Eqs (11) and (12)),
it is known that when $N=1$ the evolution described by
Eq.~(\ref{collective_langevin}) implements a process known as
quantum state exchange \cite{parkins99} which involves the
transferral of the quantum state of an electromagnetic field mode
to that of one or more trapped atoms. In particular, it is known
that, for $N=1$, Eq.~(\ref{collective_langevin}) implements this
process via the transferral of information about the input field
$a_{\rm in}$ to the ion's centre-of-mass vibrational state. It
thus follows that Eq.~(\ref{collective_langevin}) also implements
quantum state exchange when $N>1$, albeit more slowly due to an
effective decrease in $\Gamma$ with $N$ in
Eq.~(\ref{collective_langevin}).

%
%\psi_{CM}
%

In \cite{parkins2}, it was shown that we can transfer the
intracavity steady-state for the subthreshold nondegenerate
parametric amplifier $| {\rm NOPA} \rangle$ into the vibrational
states in the $x$ direction for two single trapped atoms in
different harmonic traps. Using the connection between the quantum
state exchange processes involving a single harmonically trapped
ion and $N$ harmonically trapped ions demonstrated above, it
follows that for the system illustrated in Fig.~\ref{ch3:fig_two}
we can transfer $| {\rm NOPA} \rangle$ into the centre-of-mass
modes in the $x$ direction of the two sets of $N$ trapped atoms
thus producing, in the steady state,
\begin{equation}
| \psi_{\rm CM} \rangle = \frac{1}{\cosh r}
\Sigma_{{\cal N}=0}^{\infty} \tanh^{{\cal N}} r
| {\cal N} \rangle_{1} | {\cal N} \rangle_{2},
\end{equation}
where  $| {\cal N} \rangle_{j}$ denotes the {\em centre-of-mass}
vibrational number state for the $x$ direction with eigenvalue
${\cal N}$ for the ions in the $j^{\rm th}$ trap.

For the remainder of the paper, we consider $|\psi_{\rm CM}
\rangle$ and, in Subsection~\ref{bell_scheme}, present certain
Bell inequalities involving correlations between bits in the
binary representations of ${\cal N}_{1}$ and ${\cal N}_{2}$.
Section~\ref{results} then shows that $|\psi_{\rm CM} \rangle$
violates these inequalities and thus that they highlight
nonclassical correlations in $|\psi_{\rm CM} \rangle$.

%
%what about vibrational decoherence?
%

We have assumed the trapped ions suffer no vibrational
decoherence. This assumption is justified in the following sense:
In realistic systems, the timescale over which appreciable
vibrational decoherence occurs is greater than that over which
quantum exchange would take place \cite{parkins99,parkins01}.
Because of this, we can, in principle, generate a state very
similar to $|\psi_{\rm CM} \rangle$ before appreciable vibrational
decoherence has occurred and then consider this state. Hence, even
taking vibrational decoherence into account, we can produce a
state very close to $|\psi_{\rm CM} \rangle$, thus allowing us to
ignore this decoherence of the trapped ions in our analysis.

\section{Scheme for bitwise Bell inequalities} \label{scheme}

\subsection{Motivation} \label{motivation}

To ease the reader into the Bell inequalities we consider, we now
outline a line of thinking by which someone might come to consider
the closely related bitwise Bell inequalities in \cite{larrson03}.

%
%motivation for getting to bitwise Bell inequalities
%

Each centre-of-mass vibrational number state constituting $|
\psi_{\rm CM} \rangle$ can be expressed as an infinite-length
binary string that denotes the number of centre-of-mass phonons in
the state \cite{larrson03}. For example, $| {\cal N}_{1}=2
\rangle_{1}$ can be written as $| {\cal N}_{1}= \ldots 0010
\rangle_{1}$. Expressing all centre-of-mass vibrational number
states in this manner, $| \psi_{\rm CM} \rangle$ becomes
\begin{eqnarray} \label{binary}
| \psi_{\rm CM} \rangle & = & \frac{1}{\cosh r} \left(| \ldots 000
\rangle_{1} | \ldots 000 \rangle_{2} \right. \\ \nonumber & + &
\left. \tanh r | \ldots 001\rangle_{1} | \ldots 001 \rangle_{2}
\right. \\ \nonumber & + &  \left. \tanh^{2} r| \ldots 010
\rangle_{1} | \ldots 010 \rangle_{2} \right. \\ \nonumber & + &
\left. \tanh^{3} r | \ldots 011 \rangle_{1} | \ldots 011
\rangle_{2} + \ldots \right),
\end{eqnarray}
where it is implicit that the bit strings represent ${\cal N}_{1}$ or
${\cal N}_{2}$ values.

%
%pretend that we have qubits
%

Let us for the moment pretend that each bit in Eq.~(\ref{binary})
represents a physical qubit, with corresponding bits in each
statevector with a `1'(`2') subscript representing the same qubit.
That is, with all of the least significant bits in each
statevector denoted by a `1' (`2') subscript representing one
qubit, the second least significant bits in each statevector
denoted by a `1' (`2') subscript representing another and so
forth. Upon adopting this fiction, we see that $| \psi_{\rm CM}
\rangle$ factorises as follows:
\begin{eqnarray} \label{qubit} \nonumber
| \psi_{\rm CM} \rangle & = &  \frac{1}{\cosh r} \! \left[ (| 0
\rangle_{1}^{(0)} |0 \rangle_{2}^{(0)} + \tanh r | 1
\rangle_{1}^{(0)} | 1 \rangle_{2}^{(0)}) \otimes  \right. \\
\nonumber & & \left. \bigl( | 0 \rangle_{1}^{(1)} | 0
\rangle_{2}^{(1)} + \tanh^{2} r | 1 \rangle_{1}^{(1)} | 1
\rangle_{2}^{(1)} ) \otimes
\ldots \right], \\
\end{eqnarray}
where the superscripts label pairs of qubits. Performing
single-qubit rotations on all qubits in Eq.~(\ref{qubit}) and then
making measurements in the computational basis, we can
concurrently violate the CHSH inequality for {\em all} qubit pairs
denoted by the same superscript. That is, we can simultaneously
violate this inequality for the qubit pair denoted by (0), the one
denoted by (1) and so forth.

%
%implemented approximation to ideal scheme
%

To observe $| \psi_{\rm CM} \rangle$'s  nonclassical correlations
we would like to implement the scheme involving CHSH violations
described in the previous paragraph as it produces the largest
possible violation for each qubit pair. However, our physical
system of interest does not have the distinct qubits used in the
scheme and so, in practice, we cannot address all binary digits
individually. In spite of this we can still implement a similar
scheme using other local unitaries and other measurements to
produce multiple, though smaller, CHSH violations, as shown in the
next subsection.

\subsection{The scheme} \label{bell_scheme}

%
%summarize scheme
%

In this subsection we present three CHSH inequalities involving
correlations between three pairs of bits in the binary
representations of ${\cal N}_{1}$ and ${\cal N}_{2}$. These
inequalities involve displacement operations which we apply to
both sets of ions before making certain measurements involving
electronic states.

%A measurement scheme involving electronic states \cite{dhelon96}
%is employed to measure the bits mentioned above.

%
%apply displacements
%

Applying displacements to both sets of ions in
$|\psi_{\rm CM} \rangle$ yields
\begin{equation}
| \psi_{D} \rangle = D_{1}(\alpha) D_{2}(\beta) S_{12}(r)
| 0\rangle_{1} |0 \rangle_{2},
\end{equation}
where $D_{1}(\alpha)$ and $D_{2}(\beta)$ are, respectively,
displacement operators acting on the first and second sets of ions
in $|\psi_{\rm CM}\rangle$ (that is the sets in the first and
second traps, respectively) with displacements $\alpha$ and
$\beta$. These operators are given by $D_{1}(\alpha)= \exp(\alpha
B^{(1)\;\dag}_{1x} - \alpha^{*} B^{(1)}_{1x})$ and $D_{2}(\beta)=
\exp(\beta B^{(1)\;\dag}_{2x} - \beta^{*} B^{(1)}_{2x})$. The
operator $S_{12}(r)$ is the two-mode squeezing operator which is,
when $r$ is real, $S_{12}(r)= \exp \left( r (B^{(1)\;\dag}_{1x}
B^{(1)\;\dag}_{2x} - B^{(1)}_{1x} B^{(1)}_{2x} ) \right)$, where
$r$ is a squeezing parameter. Lastly $| 0 \rangle_{1} | 0
\rangle_{2}$ is the two-mode vacuum state for the centre-of-mass
modes of the first and second sets of ions in the $x$ direction.

%
%D'Helon-Milburn measurement scheme
%

After applying $D_{1}(\alpha)$ and $D_{2}(\beta)$, the next step
in our CHSH inequality violations is to measure the values of the
$N$ least significant bits of ${\cal N}_{1}$ and ${\cal N}_{2}$.
This is done using the measurement scheme in \cite{dhelon96} which
we now describe. This scheme measures the $N$ least significant
bits of the number of centre-of-mass phonons for a set of $N$
identical two-level ions (with internal transition frequency
$\omega_{0}$) in a linear ion trap. It begins by first setting the
state of each ion to be an equal superposition of its ground and
excited internal states. Next, the measurement scheme involves
applying a standing-wave laser pulse to each ion such that each
ion's mean position coincides with a node of its pulse. The laser
frequency for all pulses $\omega_{L}$ is far detuned from all
resonant vibrational frequencies for the trapped ions and
$\Delta'$, the detuning between $\omega_{L}$ and $\omega_{0}$, is
such that $|\Delta'| \gg \nu_{x}$, where $\nu_{x}$ is the trap
frequency in the direction along which the ions are aligned. The
$m^{\rm th}$ laser pulse is applied to the corresponding ion for a
time $t_{\rm m} = 2^{m} \pi N \Delta'/(2^{N} \eta_{x}^{2}
\Omega^{2})$, where $\eta_{x}$ is the Lamb-Dickie parameter common
to all ions and $\Omega$ is the Rabi frequency for each ion.
Finally, an inverse Fourier transformation is applied to the ions'
internal states. The measurement scheme has the effect of
transferring the value of the $m^{\rm th}$ bit of the number of
centre-of-mass phonons for the ions to the two-level internal
system of the $m^{\rm th}$ ion. This bit is encoded using the
following mapping : $|g\rangle_{m} \mapsto 0$ and $|e\rangle_{m}
\mapsto 1$, where $|g\rangle_{m}$ and $|e\rangle_{m}$ are the
ground and excited internal states for the $m^{\rm th}$ ion. Once
encoded in internal states, the bit can be readily measured using
the resonant fluorescent-shelving technique \cite{heinzen90b}.

%
%show what the D'Helon-Milburn scheme produces
%

Applying the $D_{1}(\alpha)$ and $D_{2}(\beta)$ to
$|\psi_{D}\rangle$ and then using the measurement scheme on both
sets of ions in the resulting state produces
\begin{eqnarray} \nonumber
|\psi_{F} \rangle = & & \\ \nonumber
& & \hspace*{-1.40cm} \sum_{i,j=0}^{\infty,\infty} c_{i,j} | i
\rangle^{\rm vib}_{1} | j  \rangle^{\rm vib}_{2}
\otimes | {\rm binary}(  i  ,1) \rangle^{\rm e}_{1,1}
\ldots | {\rm binary}(i ,N) \rangle^{\rm e}_{1,N} \\
& & \hspace*{-1.40cm} \otimes | {\rm binary}( j ,1) \rangle^{\rm e}_{2,1} \ldots | {\rm
binary}( j ,N) \rangle^{\rm e}_{2,N},
\end{eqnarray}
where ${\rm binary}(x,y)$ is
the value of the $y^{\rm th}$ least significant binary digit of $x$ and
$c_{i,j }=\langle   i,j
| D_{1} (\alpha) D_{2}({\beta}) S_{12}(r)| 0 \rangle_{1} |0 \rangle_{2}$.
Superscript e's and vib's denote, respectively, an internal (or electronic) state
and a centre-of-mass vibrational one for the $x$ direction.
Finally, the subscript $k$ and $l$ in $|\ket_{k,l}^{\rm e}$ denote that the state is for the $l^{\rm th}$
electron in the $k^{\rm th}$ set of ions.

%
%bitwise Bell inequalities
%

We now assume, for the moment, that all observable quantities in
$|\psi_{F} \rangle$ behave classically and thus can be simulated using a local
hidden variable theory (LHVT). Given this, it follows that
the correlations between the $y^{\rm th}$ least significant bits
of ${\cal N}_{1}$ and ${\cal N}_{2}$, where $y=1,2,3 \ldots N$, can be
described by a LHVT theory.
Hence, using reasoning in \cite{clauser69},
these correlations satisfy the CHSH inequality:
\begin{eqnarray} \label{chsh_inequ}
S_{y} & = & | {\rm E} \left[ {\bar a}_{y}(\alpha) \left( {\bar
b}_{y}(\beta) + {\bar b}_{y}(\beta') \right) \right. \\ \nonumber
& & \left. + {\bar a}_{y}(\alpha') \left( {\bar b}_{y}(\beta) -
{\bar b}_{y}(\beta') \right) \right] | \leq 2,
\end{eqnarray}
where ${\bar a}_{y}(z)$ and ${\bar b}_{y}(z')$ are the values of
the $y^{\rm th}$ least significant bits of, respectively, the
first and second sets of ions given either $D_{1}(z_{1})$ or
$D_{2}(z_{2})$, where $z_{1}=\alpha,\alpha'$ and
$z_{2}=\beta,\beta'$. The notation ${\rm E}[ \ldots]$ denotes an
average or expectation value. As Inequality~(\ref{chsh_inequ})
involves thinking about ${\cal N}_{1}$ and ${\cal N}_{2}$ binary
digit by binary digit, we call the inequality in this equation a
{\em bitwise Bell inequality}. Inequality~(\ref{chsh_inequ})
arises from the fact that LHVTs are committed to the existence
definite values for all ${\bar a}_{y}$ and ${\bar b}_{y}$ at all
times that can only change in a local manner.

%
%scheme is practical
%

One important feature about the Bell-inequality scheme outlined
above is that it is potentially realistic. This is because, first,
the application of $D_{1}(z_{1})$ and $D_{2}(z_{2})$ to
$|\psi_{\rm CM} \rangle$ is feasible as existing experiments have
applied such operations to the vibrational state of a single
trapped ion (see, for example, \cite{munroe96}). Second, the
scheme is potentially realistic as the interaction between
internal and centre-of-mass vibrational states in the measurement
scheme it requires seems to be experimentally feasible. This is
the case as it only requires far-detuned standing wave laser
pulses that interact with a particular ion for set times. Finally,
it is conceivably feasible as the resonant fluorescent-shelving
technique it uses to make measurements on internal states has been
experimentally implemented with high efficiency (see, for example,
\cite{heinzen90b}).

\section{Results} \label{results}

In this section, we show that $|\psi_{\rm CM} \rangle$ violates
the three bitwise Bell inequalities represented by
Eq.~(\ref{chsh_inequ}) when $y=1,2$ or $3$. Thoughout, we assume
that $N \geq 3$ and hence that the measurement scheme can measure
up to, at least, the third least significant bits in the binary
representations of ${\cal N}_{1}$ and ${\cal N}_{2}$.

\subsection{Least significant bits} \label{least}

\begin{figure}[h]
\center{\epsfig{figure=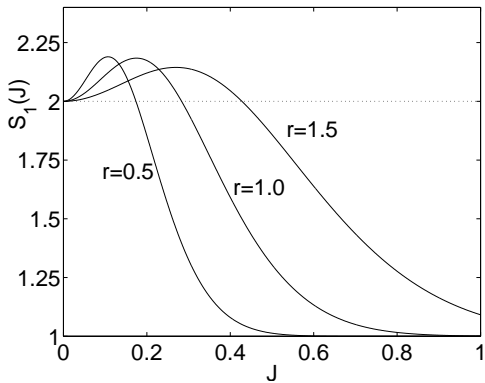,width = 65mm}}
\caption{\label{least_sig_bit_graph} Plot of $S_{1}(J)$ versus $J$
for $r=0.5$, $r=1$ and $r=1.5$ with displacements. The horizontal
dotted line represents $S_{2}(J)=2$. Both $S_{1}(J)$ and $J$ are
dimensionless.}
\end{figure}

Previous work
\cite{banasek98,banasek99a,banasek99b,chen02,larrson03,gour03} has
shown that the state $| {\rm NOPA} \ket$
%in the case where the
%subscripts `1' and `2' denote electromagnetic field modes and,
%consequently, ${\cal N}$ is the number of photons in one of these
%modes
violates instances of the CHSH inequality with the maximum
violation being $2\sqrt{2}$ \cite{chen02,larrson03,gour03}. The
violations in \cite{banasek98} were arrived at by first applying
displacement operations to modes 1 and 2 and then measuring
whether each contained an odd or even number of photons. As all
odd (even) numbers are represented by binary strings for which the
least significant bit is one (zero), \cite{banasek98}'s results
tell us that $|\psi_{\rm CM} \rangle$, which is abstractly the
same as $| {\rm NOPA} \rangle$, violates the CHSH inequality
$S_{1} \leq 2$ for the least significant bits in the binary
representations of ${\cal N}_{1}$ and ${\cal N}_{2}$ when we apply
appropriate displacement operations and then measure these bits
using the scheme in Subsection~\ref{bell_scheme}. This fact is
highlighted in Fig.~\ref{least_sig_bit_graph} which is a plot of
results formally equivalent to those in \cite{banasek98} for the
state $|\psi_{\rm CM}\ket$. In particular, for the displacements
$\alpha = \beta = 0$ and $\alpha'=-\beta' = J$, where $J \in \Re$,
it is a graph of $S_{1}$ versus $J$ for squeezing parameter values
of $r=0.5$, $r=1$ and $r=1.5$.

\subsection{Second least significant bits} \label{second_least}
In addition to violating the inequality $S_{1} \leq 2$, the state
$|\psi_{\rm CM} \rangle$ also simultaneously violates the bitwise
Bell inequality $S_{2} \leq 2$. This can be seen by calculating
the average ${\rm E} \left[ {\bar a}_{2}(z_{1}){\bar b}_{2}(z_{2})
\right]$ for this state, which is
\begin{eqnarray} \nonumber
{\rm E} \left[ {\bar a}_{2}(z_{1}) {\bar b}_{2}(z_{2}) \right] & =
& 1-2 \left( Pr({\bar a}_{2}=+1, {\bar b}_{2}=-1|z_{1},z_{2})
\right. \\ \nonumber & & \left. + Pr({\bar a}_{2}=-1, {\bar
b}_{2}=+1|z_{1},z_{2})
\right), \\
\end{eqnarray}
where $Pr({\bar a}_{2}= f, {\bar b}_{2}= g|z_{1},z_{2})$ is the
probability that ${\bar a}_{2}= f$ and ${\bar b}_{2}= g$ given the
displacements $D_{1}(z_{1})$ and $D_{2}(z_{2})$. As the second
least significant bit of a bit string is `0' for the decimal
numbers $0,1,4,5,8,9 \ldots$ and `1' otherwise,
\begin{eqnarray} \nonumber
{\rm E} \left[ {\bar a}_{2}(z_{1}) {\bar b}_{2}(z_{2}) \right]  &
= & 1 -
\\ \nonumber & & \hspace*{-2.8cm} 2 \left( \sum_{ n_{1}=0,1,4,5,
\ldots} \sum_{ n_{2} =2,3,6,7 \ldots}  Pr({\cal N}_{1}
=n_{1},{\cal N}_{2} = n_{2}|z_{1},z_{2}) \right. \\ & & \nonumber
\hspace*{-2.8cm} + \left. \sum_{n_{1} = 2,3,6,7 \ldots}
\sum_{n_{2}=0,1,4,5 \ldots} Pr( {\cal N}_{1} = n_{1}, {\cal
N}_{2}= n_{2}|z_{1},z_{2}) \right), \\
& & \label{joint_correlation}
\end{eqnarray}
where $Pr({\cal N}_{1} = n_{1},{\cal N}_{2} =n_{2}|z_{1},z_{2})$
is the probability of observing $| {\cal N}_{1} =n_{1} \rangle_{1}
| {\cal N}_{2}=n_{2} \rangle_{2}$ given the displacements
$D_{1}(z_{1})$ and $D_{2}(z_{2})$. This is known to be \cite{carl}
\begin{eqnarray} \nonumber
Pr({\cal N}_{1} = n_{1}, {\cal N}_{2} =  n_{2}|z_{1},z_{2}) & & \\
\nonumber & & \hspace*{-5.1cm} =\left| \frac{\tanh^{p} r}{\cosh r}
{\frac{p!}{q!}}^{1/2} \mu_{1}^{ n_{1}- p} \mu_{2}^{ n_{2}-q} L_{p}^{(q-p)} (\frac{ - \mu_{1} \mu_{2}}{\tanh r}) \right. \\
& & \hspace*{-5.1cm} \left. \times \exp\bigl(-(z_{1}^{*} \mu_{1} +
z_{2}^{*} \mu_{2}) / 2 \bigr) \right|^{2}, \label{j_prob}
\end{eqnarray}
where $p = {\rm min}(n_{1},n_{2})$, $q = {\rm max}(n_{1},n_{2})$,
$\mu_{1}=z_{1}+ z_{2}^{*} \tanh r$, $\mu_{2}=z_{2}+ z_{1}^{*} \tanh r$
and $L_{p}^{(q-p)}$ is a generalised Laguerre polynomial.

%
%calculate S value
%

\begin{figure}[h]
\center{\epsfig{figure=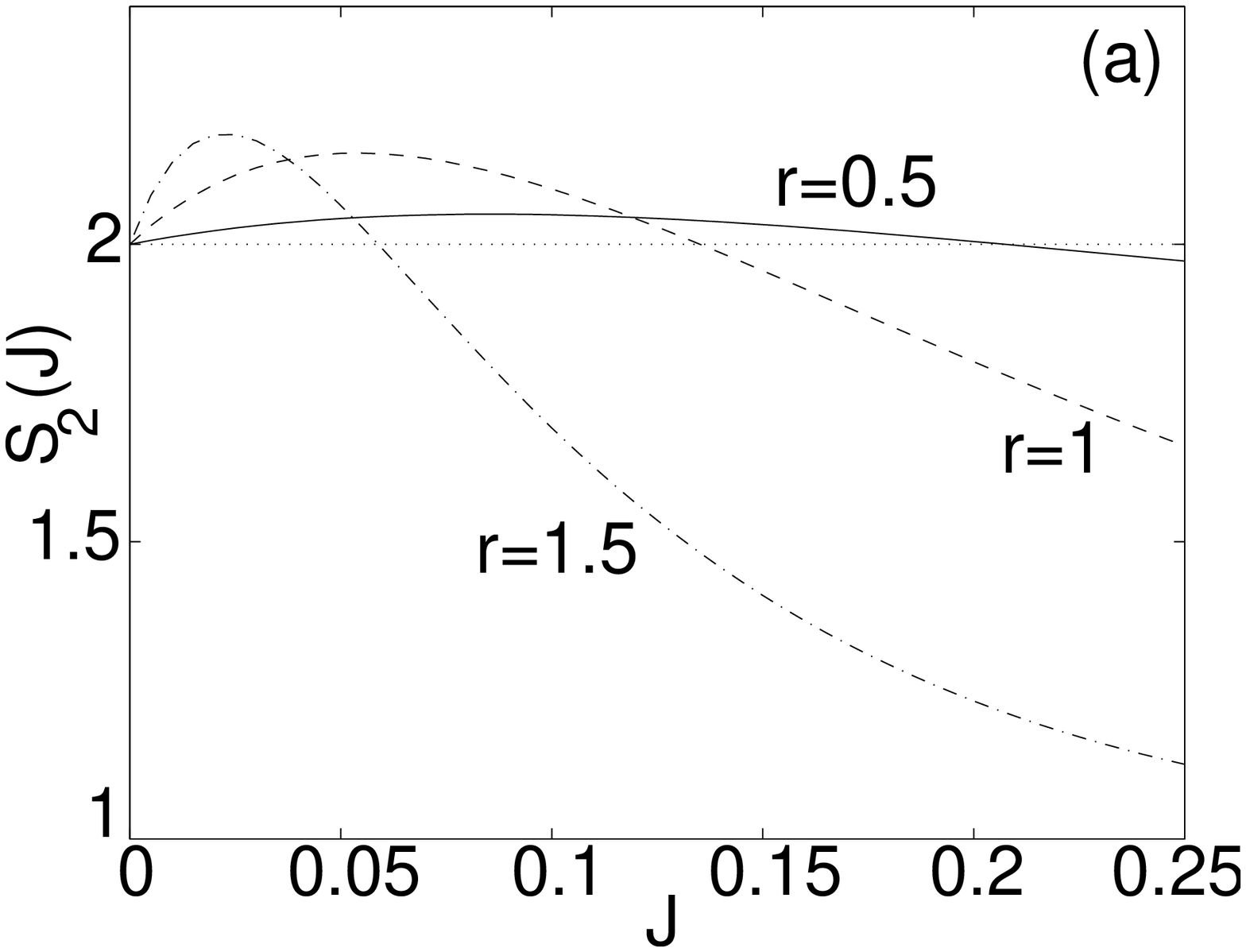,width = 65mm}}
\center{\epsfig{figure=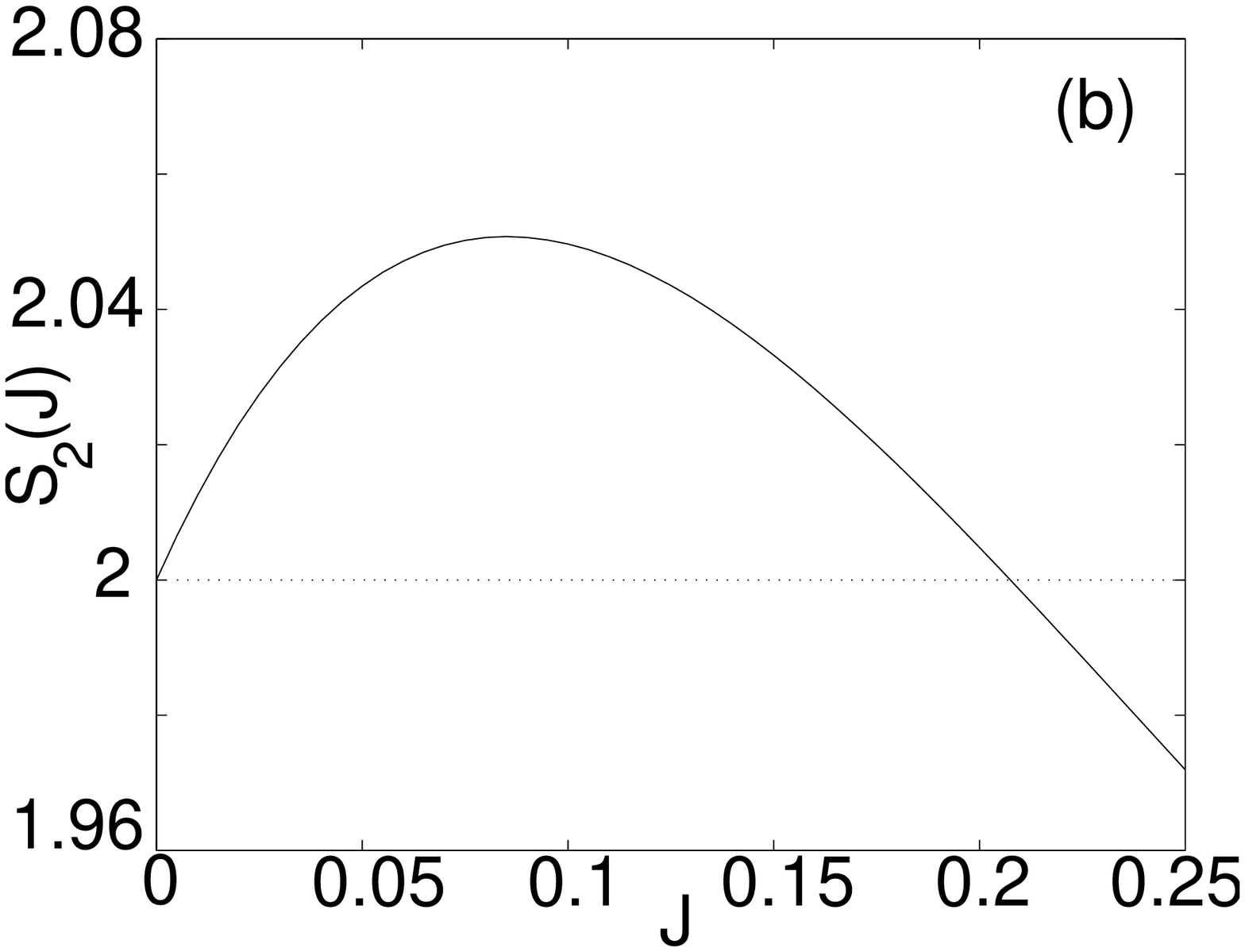,width = 65mm}}
\caption{\label{fig:second_pointfive}
$\;\;\;\;\;\;\;\;\;\;\;\;\;\;\;\;\;\;\;\;\;\;\;\;\;\;\;\;\;\;\;\;\;\;\;\;\;\;\;\;\;\;\;\;\;\;\;\;\;\;\;\;\;\;\;\;\;\;\;\;\;\;\;\;\;\;\;\;\;\;\;\;\;\;\;\;\;\;\;\;\;\;\;\;\;\;\;\;\;\;\;\;\;\;\;\;\;\;\;\;\;\;\;\;\;\;\;\;\;\;\;\;\;\;\;\;$
{\bf (a)} Plot of $S_{2}(J)$ versus $J$ for $r=0.5$ (solid line),
$r=1$ (dashed line) and $r=1.5$ (-.-.-.) with displacements. Both
$S_{2}(J)$ and $J$ are dimensionless. The horizontal dotted line
represents $S_{2}(J)=2$.
$\;\;\;\;\;\;\;\;\;\;\;\;\;\;\;\;\;\;\;\;\;\;\;\;\;\;\;\;\;\;\;\;\;\;\;\;\;\;\;\;\;\;\;\;\;\;\;\;\;\;\;\;\;\;\;\;\;\;\;\;\;\;\;\;\;\;\;\;\;\;\;\;\;\;\;\;\;\;\;\;\;\;\;\;\;\;\;\;\;$
{\bf (b)} Close-up plot of $S_{2}(J)$ versus $J$ for $r=0.5$ with
displacements. The horizontal dotted line represents
$S_{2}(J)=2$.}
\end{figure}

Calculating $S_{2}$ using Eqs~(\ref{joint_correlation}) and
(\ref{j_prob}) we obtain, upon setting $\alpha = \beta = 0$ and on
$\alpha'=-\beta' = J$, where $J \in \Re$, CHSH violations for a
range of $J$ values. These are illustrated in
Fig.~\ref{fig:second_pointfive} as a function of $J$ for squeezing
parameter values of $r=0.5$, $r=1$ and $r=1.5$. As is the case for
the graphs in Subsection~\ref{3_least} and
Section~\ref{ch4:local_squeeze}, $S_{2}$ was calculated using {\em
Mathematica}, with all numerical errors being negligible
\cite{num_errors}. Significantly, for a range of $J$ and $r$
values we {\em simultaneously} violate the bitwise Bell
inequalities for the least and second least significant bits in
the binary representations of ${\cal N}_{1}$ and ${\cal N}_{2}$,
as can be determined by inspecting Fig.~\ref{fig:second_pointfive}
and results in \cite{banasek98}.

\subsection{Third least significant bits} \label{3_least}

%
%displacement and third least significant digit.
%

\begin{figure}
\center{\epsfig{figure=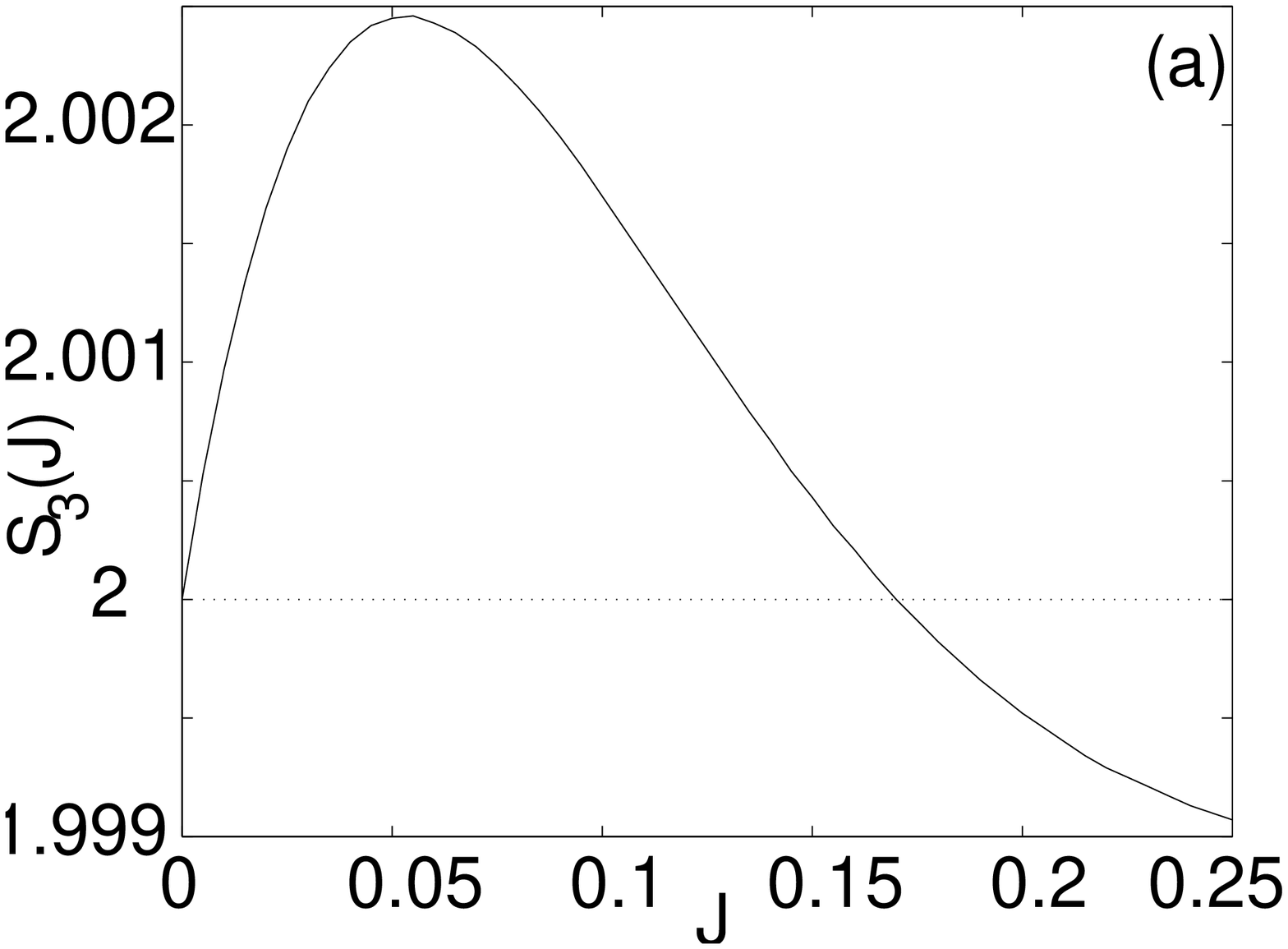,width=65mm}}
\center{\epsfig{figure=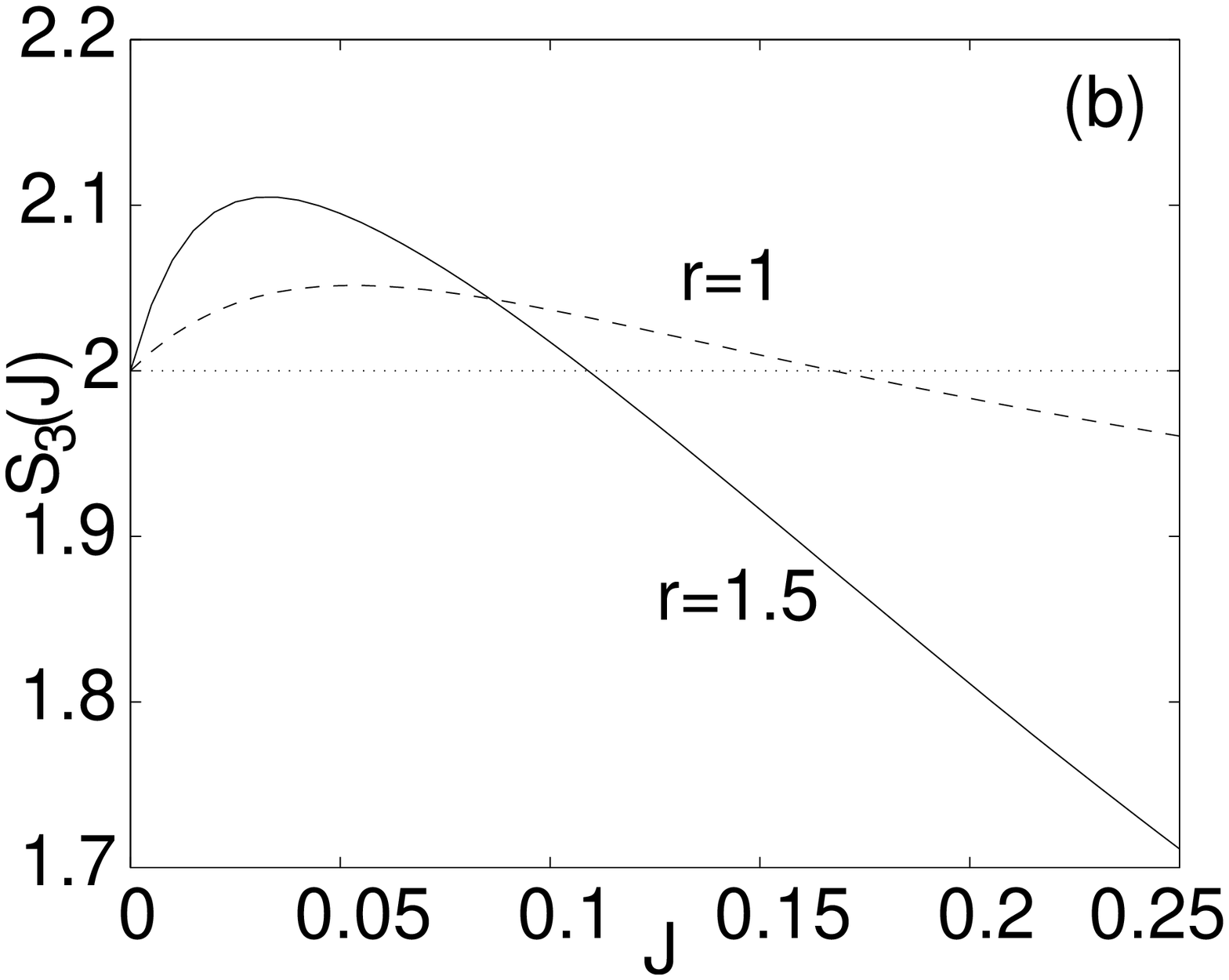,width=65mm}}
\caption{\label{ch4:fig3}
$\;\;\;\;\;\;\;\;\;\;\;\;\;\;\;\;\;\;\;\;\;\;\;\;\;\;\;\;\;\;\;\;\;\;\;\;\;\;\;\;\;\;\;\;\;\;\;\;\;\;\;\;\;\;\;\;\;\;\;\;\;\;\;\;\;\;\;\;\;\;\;\;\;\;\;\;\;\;\;\;\;\;\;\;\;\;\;\;\;\;\;\;\;\;\;\;\;\;\;\;\;\;\;\;\;\;\;\;\;\;\;\;\;\;\;\;\;\;\;\;\;$
{\bf (a)} Plot of $S_{3}(J)$ versus $J$ for $r=0.5$ with
displacements. Both $S_{3}(J)$ and $J$ are dimensionless. The
horizontal dotted line represents $S_{3}(J)=2$.
$\;\;\;\;\;\;\;\;\;\;\;\;\;\;\;\;\;\;\;\;\;\;\;\;\;\;\;\;\;\;\;\;\;\;\;\;\;\;\;\;\;\;\;\;\;\;\;\;\;\;\;\;\;\;\;\;\;\;\;\;\;\;\;\;\;\;\;\;\;\;\;$
{\bf (b)} Plot of $S_{3}(J)$ versus $J$ for $r=1.5$ (solid line)
and $r=1$ (dashed line) with displacements. The horizontal dotted
line represents $S_{3}(J)=2$.}
\end{figure}

To show that $|\psi_{\rm CM}\rangle$ violates the bitwise Bell
inequality $S_{3} \leq 2$, we now perform a similar calculation to
the last subsection's except that, as the third least significant
bit of the numbers $0,1,2,3,8,9,10,11 \ldots$ is `0',
\begin{eqnarray} \label{three_expectation} \nonumber
{\rm E} \left[ {\bar a}_{3}(z_{1}) {\bar b}_{3}(z_{2}) \right] & =
1- 2\times &
\\ \nonumber
& & \hspace*{-4.0cm}\left( \sum_{i = {\rm even}} \sum_{j=0}^{3}
\sum_{l ={\rm odd}} \sum_{s=0}^{3} Pr({\cal N}_{1}=4i + j,{\cal
N}_{2}=4l+s|z_{1},z_{2}) \right. \\ \nonumber
& & \hspace*{-4.0cm}
\left.  \sum_{i={\rm odd}} \sum_{j=0}^{3} \sum_{l={\rm even}}
\sum_{s=0}^{3}  Pr({\cal N}_{1}=4i+j, {\cal
N}_{2}=4l+s|z_{1},z_{2}) \right). \\
\end{eqnarray}
Using this result and Eq.~(\ref{j_prob}) to calculate $S_{3}$ as a
function of $J$, we obtain the results in Figs~\ref{ch4:fig3} (a)
and (b) which show bitwise Bell inequality violations. Observe
that for certain $J$ and $r$ values, we can simultaneously violate
the bitwise Bell inequalities $S_{1} \leq 2$, $S_{2} \leq 2$ and
$S_{3} \leq 2$. It is also interesting to note that the violations
in Fig.~\ref{ch4:fig3} are significantly less than those for the
second least significant bits shown in
Fig.~\ref{fig:second_pointfive}. It is possible that this is due
to the fact that groups of four consecutive numbers (eg. 0, 1, 2
and 3) share the same value for their third least significant
bits. (For the second least significant bits, only two consecutive
numbers share the same value.) Because of this, it may be more
difficult for the displacement operations we implement to cause
states to `flip' the values of their third least significant bits.
In turn, this would mean that it would be more difficult for these
operations to generate the sort of interference between previously
orthogonal states in $|\psi_{\rm CM} \ket$ necessary for obtaining
Bell violations, thus leading to the smaller violations for the
third least significant bits shown in Fig~\ref{ch4:fig3}.

\section{Bitwise Bell violation with local squeezing operations} \label{ch4:local_squeeze}

%
%look at other things than squeezing
%

In this subsection we show that unitaries other than displacement
operations can yield a bitwise Bell inequality violation for
$|\psi_{\rm CM} \rangle$. In particular, we show that squeezing
operations applied to the centre-of-mass vibrational states of
both sets of ions in the $x$ direction can produce such a
violation involving the second least significant bits of ${\cal
N}_{1}$ and ${\cal N}_{2}$. These squeezing operations are
interesting to consider as they have been practically implemented
in ion traps (see, for example, \cite{meekhof96}). Observe,
however, that squeezing operations applied to both sets of ions do
not produce CHSH inequality violations for the least significant
bits in the the binary representations of ${\cal N}_{1}$ and
${\cal N}_{2}$ as squeezing operations are associated with
two-phonon creation and annihilation. Thus, they do not cause odd
and even phonon number states to change parity and so do not
induce the type of interference required for such violations.
Throughout this section we assume that $N \geq 2$, so that the
measurement scheme involves enough two-level electronic systems to
to measure ${\bar a}_{2}$ and ${\bar b}_{2}$.

%
%calculations for squeezing
%

Applying the above mentioned squeezing operations to $|\psi_{\rm
CM} \rangle$, we obtain
\begin{equation}
| \psi_{S} \rangle = S_{1}(r_{+})S_{2}(r_{-})S_{12}(r)
| 0,0 \rangle,
\end{equation}
where $S_{1}$ and $S_{2}$ are single-mode squeezing operators for
the centre-of-mass modes of the first and second sets of ions in
$|\psi_{\rm CM} \rangle$ in the $x$ direction with real squeezing
parameters $r_{+}$ and $r_{-}$. The operator
$S_{1}(r_{+})=\exp\left(r_{+} \left( B_{1x}^{(1)\;\dag}\right)^{2}
- r_{+ } \left( B_{1x}^{(1)} \right)^{2} \right)$, whilst
$S_{2}(r_{-})=\exp\left(r_{-} \left( B_{2x}^{(1)\;\dag}\right)^{2}
- r_{-} \left( B_{2x}^{(1)}\right)^{2}      \right)$. We now
determine $Pr_{\rm squeeze}({\cal N}_{1}=n_{1},{\cal
N}_{2}=n_{2}|r_{+},r_{-})$, the probability of observing $n_{1}$
and $n_{2}$ centre-of-mass phonons in the $x$ direction for the
first and second sets of ions respectively given $S_{1}(r_{+})$
and $S_{2}(r_{-})$ by re-ordering the operators in
$S_{1}(r_{+})S_{2}(r_{-})S_{12}(r)|0 \rangle_{1}0\rangle_{2}$. The
idea for this derived from a calculation in \cite{carl} that found
$\langle {\cal N}_{1}=n_{1}, {\cal N}_{2}=n_{2} | \psi_{D}
\rangle$ by decomposing and normally ordering the operators in
$D_{1}(\alpha) D_{2}(\beta) S_{12}(r) |0 \rangle_{1}
0\rangle_{2}$.

Decomposing $S_{12}$ in a normally-ordered manner and utilizing
the known single-mode squeezing operator decomposition
\cite{schumaker}
\begin{eqnarray} \nonumber
S_{j}(R) &=&\exp \bigl[ -{\rm ln}(\cosh R)( B_{jx}^{(1)\;\dag}
B_{jx}^{(1)} +1/2) \bigr] \times \\ \nonumber & & \exp \bigl[
\tanh R \cosh^{2} R \left( B_{jx}^{(1)\;\dag} \right)^{2}  /2
\bigr] \times \\
& & \exp \bigl[ \tanh R  \left(  B_{jx}^{(1)}  \right)^{2}
/2 \bigr],
\end{eqnarray}
where $j=1,2$, yields
\begin{eqnarray} \label{normal} \nonumber
| \psi_{S} \rangle & = & \frac{1}{\sqrt{K}}  \times \\ \nonumber &
& \hspace*{-1.4cm} \exp\left( - \ln \cosh_{r+} B_{1x}^{(1)\;\dag}
B_{1x}^{(1)} - \ln \cosh_{r-} B_{2x}^{(1)\;\dag} B_{2x}^{(1)} \right) \times \\
\nonumber
& & \hspace*{-1.4cm} \exp \left( -\tanh r_{+} \cosh^{2} r_{+}
\left( B_{1x}^{(1)\;\dag} \right)^{2}/2 \right. \\ \nonumber & &
\hspace*{-0.75cm} \left. - \tanh r_{-}
\cosh^{2} r_{-}  \left( B_{2x}^{(1)\;\dag} \right)^{2}/2 \right) \times \\
\nonumber
& & \hspace*{-1.4cm} \exp \left( \tanh r_{+} \left(
B_{1x}^{(1)} \right)^{2} /2
+ \tanh r_{-} \left(B_{2x}^{(1)} \right)^{2} /2 \right) \times \\
& & \hspace*{-1.4cm} \exp(\tanh r B_{1x}^{(1)\;\dag}
B_{2x}^{(1)\;\dag}) | 0,0 \rangle,
\end{eqnarray}
where $K = \cosh^{2} r \cosh r_{+} \cosh r_{-}$.

%
%doing maths to find P_{squeeze}(n_1,n_2)
%

To determine $Pr_{\rm squeeze}({\cal N}_{1}=n_{1},{\cal
N}_{2}=n_{2}|r_{+},r_{-})$ from Eq.~(\ref{normal}) we use three
operator identities. These identities, which hold for any
operators $A$ and $B$ such that $[A,A^{\dag}] = [B,B^{\dag}]=1$
and $[A,B^{\dag}] = [A,B]=0$, are proven in Appendix A and are
\begin{eqnarray} \label{lie_identities} \nonumber
\exp(c_{1} A^{2}) \exp(c_{2} A^{\dag} B^{\dag}) & = &
\exp(c_{2} A^{\dag} B^{\dag})
\exp(c_{1}c_{2}^{2}B^{\dag\:2}) \\
\nonumber
& \times & \exp(2 c_{1} c_{2} A B^{\dag})\; \exp(c_{1} A^{2}), \\
& & \\
\nonumber & & \hspace*{-4.3cm} \exp(c_{1}A^{2})\exp(c_{2}
A^{\dag\:2}) =
\exp(\frac{c_{2}}{1 - 4c_{1} c_{2}} A^{\dag\:2}) \times \\
& & \hspace*{-4.1cm} \exp\bigl[ \cosh^{-1}\bigl(  1  + \frac{2
c_{1}c_{2}}{1-4c_{1}c_{2}} -2c_{1}c_{2}\bigr) \{A^{\dag}A +1/2\}
\bigr] \times \nonumber \\
& & \hspace*{-4.1cm} \exp(\frac{ c_{1}}{1 - 4 c_{1} c_{2}} A^{2})
\\ \nonumber
& & {\rm and} \\
\nonumber & & \hspace*{-4.1cm} \exp(c_{1} A^{\dag} B) \exp(c_{2}
B^{\dag\:2})  = \exp(c_{2} B^{\dag\:2}) \; \exp(2 c_{1} c_{2}
A^{\dag} B^{\dag})  \\
\nonumber & & \times \exp(c_{1}^{2} c_{2}
A^{\dag\:2})\; \exp(c_{1} A^{\dag} B), \\
& &
\end{eqnarray}
where $c_{1}$ and $c_{2}$ are complex numbers. Re-ordering terms
on the right-hand side of Eq.~(\ref{normal}) by applying these
identities to $B_{1x}^{(1)}$ and $B_{2x}^{(1)}$ yields
\begin{eqnarray} \label{squeeze_projection}
| \psi_{S} \rangle & = & \sqrt{\frac{M}{K}} \times
\\ \nonumber
& & \hspace*{-1.3cm} \exp
\left(-\ln \cosh r_{+} B_{1x}^{(1)\;\dag} B_{1x}^{(1)} - \ln \cosh
r_{-} B_{2x}^{(1)\;\dag} B_{2x}^{(1)} \right) \times \\ \nonumber & &
\hspace*{-1.3cm} \exp \left[ ( - 1/2 \tanh r_{+} \cosh^{2} r_{+} +
1/2 \tanh r_{-} \tanh^{2} r \right. \\ \nonumber & & \left. +
d_{1}^{2} d_{2}) \left( B_{1x}^{(1)\;\dag} \right)^{2} \right] \times \\
\nonumber
& & \hspace*{-1.3cm} \exp \left[ \left( -1/2 \tanh r_{-} \cosh^{2} r_{-}
+ d_{2}) \left( B_{2x}^{(1)\;\dag} \right)^{2} \right) \right. \times \\
\nonumber
& & \hspace*{-1.3cm} \left. \exp \left( (\tanh r + 2 d_{1} d_{2})
B_{1x}^{(1)\;\dag} B_{2x}^{(1)\;\dag} \right) \right] |0, 0
\rangle,
\end{eqnarray}
where $d_{1} =\tanh r \tanh r_{-}$,
$d_{2} = d_{4}/1-4d_{3}d_{4}$,
$d_{3}=1/2\tanh r_{-}$,
$d_{4}=1/2 \tanh r_{+} \tanh^{2} r$
and
\begin{displaymath}
M  = \exp \bigl[ \cosh^{-1} \bigl( 1 + \frac{2 d_{3} d_{4}}
{1- d_{3} d_{4}} - 2 d_{3} d_{4} \bigr) \bigr].
\end{displaymath}
\

%
%calculate P(n_1,n_2)
%

\begin{figure}
\center{\epsfig{figure=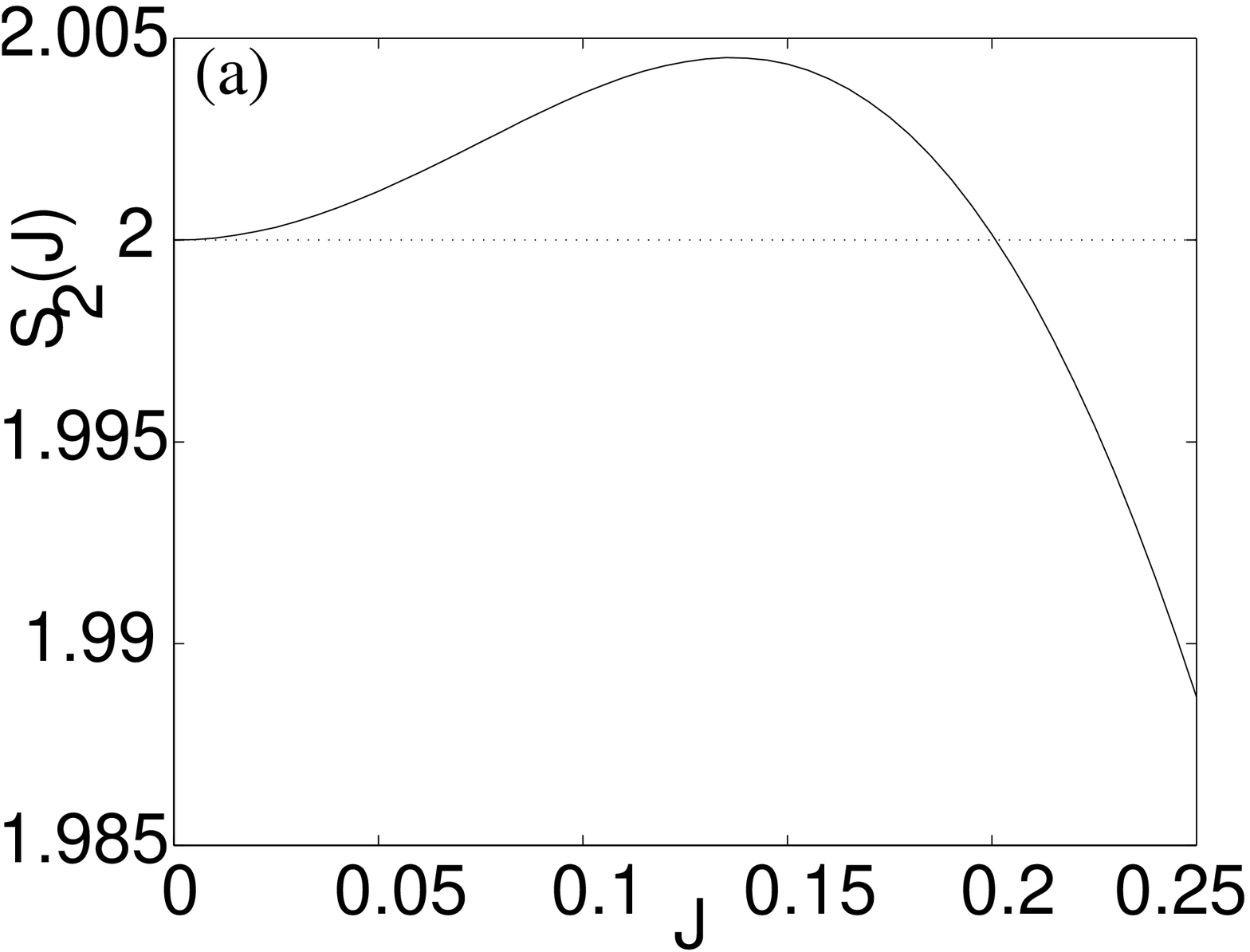,width=65mm}}
\center{\epsfig{figure=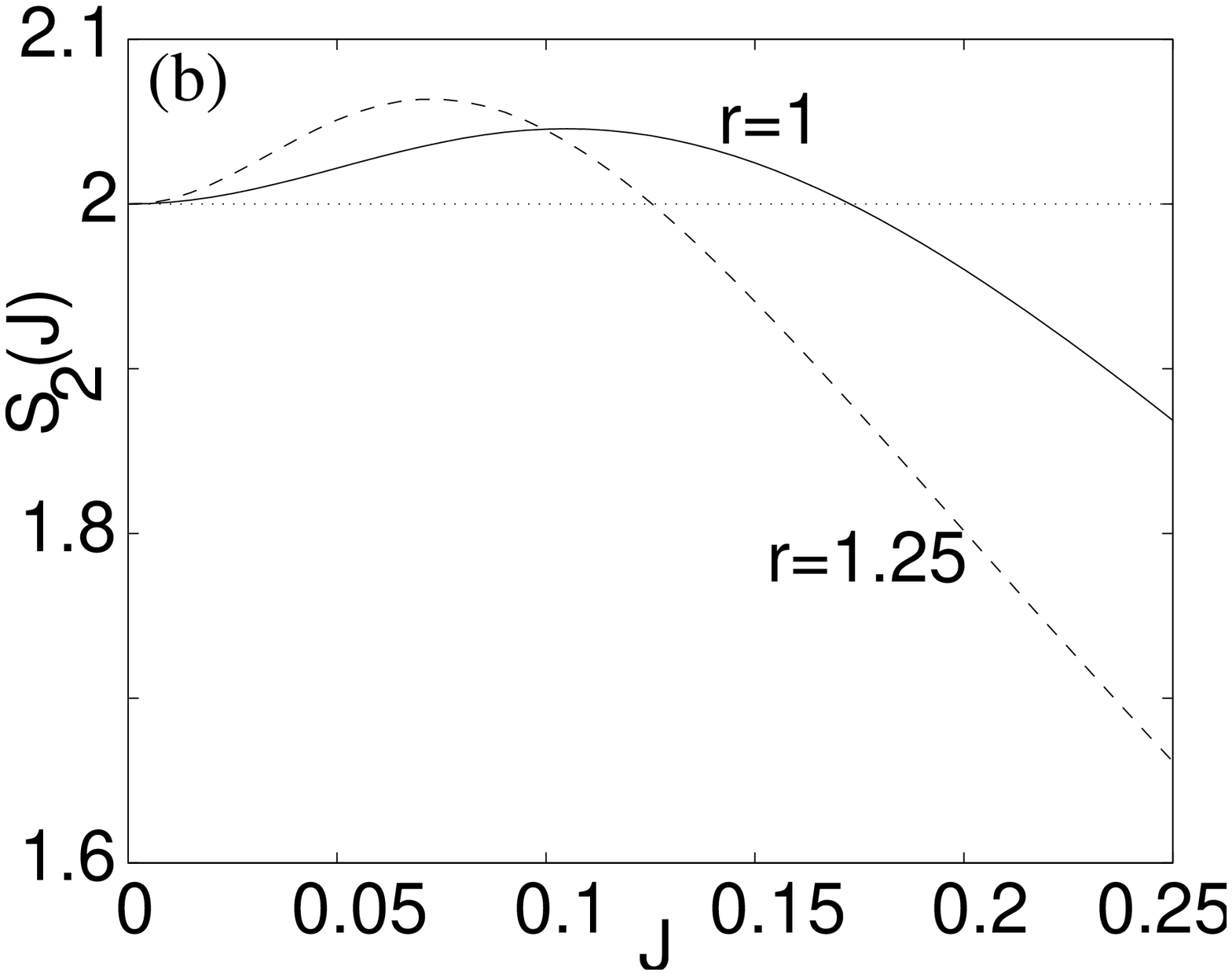,width=65mm}}
\caption{\label{ch4:squeeze_one}\newline(a) Plot of $S_{2}(J)$
versus $J$ for $r=0.5$ with local squeezing operations. Both
$S_{2}(J)$ and $J$ are dimensionless. The horizontal dotted line
represents the equation $S_{2}(J)=2$.
\newline (b) Plot of $S_{2}(J)$ versus $J$ for $r=1$ (solid line) and $r=1.25$ (dashed line) with local
squeezing operations. The horizontal dotted line represents the
equation $S_{2}(J)=2$.}
\end{figure}

Calculating $Pr_{\rm squeeze}({\cal N}_{1}=n_{1},{\cal
N}_{2}=n_{2}|r_{+},r_{-})$ using the right-hand side of
Eq.~(\ref{squeeze_projection}), we arrive at
\begin{eqnarray} \nonumber
Pr_{\rm squeeze}({\cal N}_{1}=  n_{1},{\cal
N}_{2}=n_{2}|r_{+},r_{-}) & = & f(n_{1},n_{2}) \times \\ \nonumber
& & \hspace*{-6cm} \left| \sqrt{\frac{M}{K}} \sqrt{n_{1}!n_{2}!}
\cosh^{-n_{1}} r_{+}
\cosh^{-n_{2}} r_{-} e_{+}^{n_{1}/2} e_{-}^{n_{2}/2} \right. \\
\nonumber & & \hspace*{-6cm} \left. \sum_{j=0+f,2+f,4+f
\ldots}^{{\rm min}(n_{1},n_{2})} (\frac{e_{\rm two}}{\sqrt{e_{+}
e_{-}}})^{j} \frac{1}{j!((n_{1}-j)/2)!((n_{2}-j)/2)!} \right|^{2},
\\
& & \label{final}
\end{eqnarray}
where
\begin{equation}
f(n_{1},n_{2}) = \left\{
\begin{array}{cc}
0 & , n_{1}+n_{2}=\:{\rm even} \\
1 & , n_{1}+n_{2}=\:{\rm odd},
\end{array}
\right.
\end{equation}
$e_{+}= - 1/2 \tanh r_{+} \cosh^{2} r_{+} + 1/2 \tanh r_{-}
\tanh^{2} r + d_{1}^{2} d_{2}$, $e_{-}=-1/2 \tanh r_{-} \cosh^{2}
r_{-} + d_{2}$ and $e_{\rm two}=\tanh r + 2 d_{1} d_{2}$. Observe
that it was crucial to re-order the operators on the right-hand
side of Eq.~(\ref{normal}) in arriving at Eq.~(\ref{final}) as if
we did not then we would have had to deal with infinitely many
terms contributing to $P_{\rm squeeze}({\cal N}_{1}=n_{1},{\cal
N}_{2}=n_{2}|r_{+},r_{-})$. The reason for this is that the
right-hand side of Eq.~(\ref{normal}) contains annihilation
operators to the left of creation operators for the same mode. As
a consequence, for example, upon considering power series
expansions of the exponentials in this equation we have
contributions to $|0,0\rangle$ from terms in which we first create
$X$ centre-of-mass phonons, where $X=1,2,3 \ldots$, by applying
$\left( B_{1x}^{(1)\dag} \right)^{X}$, and then annihilate them by
applying $\left( B_{1x}^{(1)} \right)^{X}$ to $\left(
B_{1(x)}^{(1)\;\dag} \right)^{X} |0,0\rangle$. Given that $X$ can
be any natural number, it follows that, to determine $Pr_{\rm
squeeze}({\cal N}_{1}=0, {\cal N}_{2}=0|r_{+},r_{-})$ using
Eq.~(\ref{normal}), we seem to need to consider infinitely many
contributing terms. In contrast, the only annihilation operators
that appear to the left of creation operators in the right-hand
side of Eq.~(\ref{squeeze_projection}) are present in terms
containing number operators. These do not increase or decrease the
number of centre-of-mass phonons when applied to a state and so
their presence does not cause infinitely many terms to contribute
to $P_{\rm squeeze}({\cal N}_{1}=n_{1},{\cal
N}_{2}=n_{2}|r_{+},r_{-})$, thus making the calculation of $P_{\rm
squeeze}({\cal N}_{1}=n_{1},{\cal N}_{2}=n_{2}|r_{+},r_{-})$
tractable. Using $Pr_{\rm squeeze}({\cal N}_{1}=n_{1},{\cal
N}_{2}=n_{2}|r_{+},r_{-})$ to calculate CHSH correlations in a
similar manner to that used to determine $Pr({\cal
N}_{1}=n_{1},{\cal N}_{2}=n_{2}|z_{1},z_{2})$ in
Subsection~\ref{second_least}, we find that the CHSH inequality is
violated, as illustrated in Figs~\ref{ch4:squeeze_one} (a) and (b)
which show $S_{2}$ as a function of $J$ for $r=0.5$, $r=1$ and
$r=1.25$.

\section{Discussion} \label{discussion}

%
%how this stuff differs from Larrson's paper
%

Though this paper's results are related to those in
\cite{larrson03}, they differ from \cite{larrson03}'s results in a
number of ways. First, motivated by the comment in
\cite{larrson03} that the formulation of a more practical
measurement scheme to measure \cite{larrson03}'s Bell inequalities
was desirable, we arguably proposed such a scheme (at least for
the least, second least and third least significant bits of ${\cal
N}_{1}$ and ${\cal N}_{2}$) which centred on transferring the
centre-of-mass vibrational state of group of trapped ions to the
internal states of a group of electrons. Second, we violated
inequalities similar to those in \cite{larrson03} using vastly
different schemes to those in \cite{larrson03}. In
\cite{larrson03} the measurements made in violating the
inequalities were pseudo-spin measurements along varying axes in a
two-dimensional plane. In contrast, we measured a pseudo-spin
based observable in a single direction and obtain violations by
applying a range of displacements and squeezing operations to sets
of $N$ ions.

%
%why this paper is interesting
%

Aside from extending work in \cite{larrson03}, the Bell violations
in Section~\ref{results} are noteworthy as they are similar to
those attainable in so-called hyper-entangled states
\cite{kwiat97}. These are states in which more than four degrees
of freedom are entangled, such as a two-photon state with
polarisation, energy and momentum entanglement (each particle has
a polarisation, an energy and a momentum degree of freedom
participating in the entanglement). As a consequence,
hyper-entangled states can violate multiple Bell inequalities
involving, collectively, five or more degrees of freedom. The Bell
violations in Section~\ref{results} are similar to those
achievable in hyper-entangled states as the violations in
Section~\ref{results} involve violating three Bell inequalities
involving six degrees of freedom, namely the three least
significant bits of both ${\cal N}_{1}$ and ${\cal N}_{2}$. One
reason why this connection is interesting is that indirect
evidence suggests \cite{kwiat97} (pp. 2179-81) that some
hyper-entangled states may be able to perform certain interesting
quantum information processing. This, in turn, suggests that
$|\psi_{\rm CM} \rangle$ may also be able to perform such feats.

%
%be cautious about multiple degrees of freedom in violations for \psi_{\rm CM}
%

%Though the Bell violations in Section~\ref{results} are similar to
%those achievable in hyperentangled states, in some ways they are
%less interesting. First, the eigenvectors of half of these
%degrees, the three least significant bits of ${\cal N}_{i}$, for
%$i=1,2$, lie in the same Hilbert space, namely that for the
%$i^{\rm th}$ set of $N$ trapped ions. In contrast, at most two
%degrees of freedom involved in attainable Bell violations for the
%hyperentangled state in \cite{kwiat97} reside in the same Hilbert
%space. Another difference between the two Bell violations is that
%the violations in Section~\ref{results} involve local operators,
%namely local displacements, that simultaneously act on half of the
%degrees of freedom, as opposed to just one of the them. The two
%above features of the Bell violations in Section~\ref{results}
%suggest that these violations are connected to each other and thus
%not as interesting as the more independent violations achievable
%in the hyper-entangled states in \cite{kwiat97}.

%
%4th, 5th ... bits
%

Though we only demonstrated bitwise Bell violations for the three
least significant bits of ${\cal N}_{1}$ and ${\cal N}_{2}$
others, presumably, also exist for the fourth, fifth, sixth etc.
least significant bits. However, the calculations required to
demonstrate these violations were not performed as calculating
$S_{y}$ becomes increasingly difficult as $y$ increases due to the
presence of more and more complicated spreads of centre-of-mass
number states sharing the same value for the $y^{\rm th}$ bit. An
example of this increased complication can be seen by observing
the fact that Eq.~(\ref{joint_correlation}) ($y=2$) is simpler
than Eq.~(\ref{three_expectation}) ($y=3$).

%
%more qip power.
%

An alternate approach we could have taken to investigating
$|\psi_{\rm CM}\rangle$'s nonclassical correlations would have
been to see what quantum information processing tasks this state's
correlations could be used to perform. However, one complication
with this is that the $2N$ systems in the total physical system
described by $|\psi_{\rm CM}\rangle$ are not qubits but instead
are infinite-dimensional harmonic oscillators. Because of this, we
cannot directly consider if this state is useful in helping to
implement well-known quantum protocols for qubits. In spite of
this difficulty, however, a recent result showing that certain
Bell violations imply the existence of quantum communication
complexity protocols superior to any classical ones
\cite{brukner02} may be useful in manifesting $| \psi_{\rm
CM}\rangle$'s nonclassical correlations. In particular, it may
allow us to readily show that Section~\ref{results}'s Bell
violations imply that $|\psi_{\rm CM}\rangle$ could be employed to
perform such quantum protocols.

%
%entanglement witnesses
%

Yet another approach that could be taken to illustrate the
nonclassicality of $| \psi_{\rm CM} \rangle$ is to use
entanglement witnesses \cite{terhal00}. An entanglement witness $W$ for the entangled state
$\rho$ is an operator such that $Tr(\rho W) < 0$ and $Tr(\sigma W)
> 0$ whenever $\sigma$ is a separable state.
This approach would involve identifying suitable operators $W$ and
then applying them to $| \psi_{\rm CM} \rangle$. We acknowledge
that it may be a useful approach to try, however, we have not
explored it.

%
%practicality?
%

How feasible are the system and measurement scheme we have
discussed? To reiterate, first, the system involving the
parametric oscillators feeding into two cavities within which lie
ion traps containing $N$ ions does not seem to be infeasible. This
is because, as stated earlier, optical cavities and nondegenerate
optical parametric amplifiers have been widely realized in
laboratories for some time. In addition, experiments in which a
single harmonically trapped ion has been placed within an optical
cavity have been conducted \cite{mundt02}. Another factor
consistent with the potential feasibility of the system considered
is that the entangled state $|\psi_{\rm CM }\rangle$ can be
created, to a good approximation, on a timescale far shorter than
that of the vibrational decoherence for the ions. Finally,
displacement and squeezing operations on trapped ions have been
realised in \cite{meekhof96} via shining laser beams on the ions.

%
%conclusion
%

To conclude, following on from \cite{larrson03} we have presented
Bell inequalities that reveal certain nonclassical correlations in
$|\psi_{\rm CM}\rangle$. In particular, these correlations are
between bits in the binary representations of ${\cal N}_{1}$ and
${\cal N}_{2}$ that violate three bitwise inequalites. We have
also presented a bitwise Bell violation for $|\psi_{\rm
CM}\rangle$ involving local squeezing operations.

\section*{Acknowledgements}
Both authors wish to thank Professor S. Braunstein for stimulating
discussions. DTP thanks Professor T. Bracken and Dr J. Links for
assistance with Lie algebras and also acknowledges the assistance
of H. M. Wolffram.

\section*{Appendix A: Operator identities using Lie algebras} \label{app_a}

%\appendix{Appendix A: Operator identities using Lie algebras} \label{app_a}

%
%Lie algebra operator identity for two mode squeezed state
%

In this appendix we prove the identities
\begin{eqnarray} \nonumber
\exp(c_{1} A^{2}) \exp(c_{2} A^{\dag} B^{\dag}) & = & \exp(c_{2}
A^{\dag} B^{\dag})
\exp(c_{1}c_{2}^{2}B^{\dag\:2}) \\
\nonumber
& \times & \exp(2 c_{1} c_{2} A B^{\dag})\; \exp(c_{1} A^{2}), \\
\label{identity1} & & \\
\nonumber & & \hspace*{-4.3cm} \exp(c_{1}A^{2})\exp(c_{2}
A^{\dag\:2}) =
\exp(\frac{c_{2}}{1 - 4c_{1} c_{2}} A^{\dag\:2}) \times \\
& & \hspace*{-4.1cm} \exp\bigl[ \cosh^{-1}\bigl(  1  + \frac{2
c_{1}c_{2}}{1-4c_{1}c_{2}} -2c_{1}c_{2}\bigr) \{A^{\dag}A +1/2\}
\bigr] \times \nonumber \\
\label{identity2} & & \hspace*{-4.1cm}
\exp(\frac{ c_{1}}{1 - 4 c_{1} c_{2}} A^{2})
\\ \nonumber
& & {\rm and} \\
\nonumber & & \hspace*{-4.1cm} \exp(c_{1} A^{\dag} B) \exp(c_{2}
B^{\dag\:2})  = \exp(c_{2} B^{\dag\:2}) \; \exp(2 c_{1} c_{2}
A^{\dag} B^{\dag})  \\
\nonumber & & \times \exp(c_{1}^{2} c_{2}
A^{\dag\:2})\; \exp(c_{1} A^{\dag} B), \\
\label{identity3} & &
\end{eqnarray}
where the $c_{1}$ and $c_{2}$ are $c$-number co-efficients and $A$
and $B$ are bosonic centre-of-mass annihilation operators for two
non-interacting systems for which $[A,A^{\dag}] = [B,B^{\dag}] =
1$. We do this using the Baker-Campbell-Hausdorff (BCH) formula
\cite{varadarajan84}, a technique that has been called the {\em
differential-equation approach} \cite{schumaker,truax88} and the
fact that SU(1,1) has a two-dimensional matrix representation.

%
%explain BCH formula and the d.e. approach
%

The BCH formula is \cite{varadarajan84} (p. 118)
\begin{eqnarray}
\exp({\cal A}) \exp( {\cal B}) & =  & \\ \nonumber & &
\hspace*{-2.9cm} \exp \bigl( {\cal A} + {\cal B} + 1/2 [ {\cal
A},{\cal B}] + 1/12 \{ [ {\cal
A},[{\cal A},{\cal B}]] + [[{\cal A}, {\cal B}],{\cal B}] \} \\
\nonumber & & \hspace*{-2.9cm} + 1/48 \{ [ {\cal B},[ {\cal
A},[{\cal B},{\cal A}]]] + [[[{\cal A},{\cal B}],{\cal A}],{\cal
B}] \} + \ldots \bigr),
\end{eqnarray}
where ${\cal A}$ and ${\cal B}$ are abitrary operators. We use it
in proving Eqs~(\ref{identity1}) and (\ref{identity3}) by
employing it to convert their left-hand sides into a single
exponential each. Next, we convert these single exponentials to
the normally-ordered products of exponentials on the right-hand
sides of Eqs~(\ref{identity1}) and (\ref{identity3}) using the
differential-equation approach, which we now explain. This
approach disentangles or decomposes a single exponential with a
sum in its exponent, into a product of a number of exponentials in
the following manner: First, we multiply the exponent of the
single exponential by a parameter $t$. Next, we equate the single
exponential with this extra factor of $t$ in its exponent to a
product of exponentials for which each exponent is some unknown
function of $t$ multiplied by a generator of a certain Lie group.
The group is the same for all exponents and is one for which all
terms in the exponent of the original single exponential are some
constant multiplied by a generator of the group. In addition, each
generator appears precisely once in the product of exponentials.
To give an example, consider the single exponential
$\exp(c_{1}A^{\dag\:2} + c_{2} A^{2})$. Noting that $A^{2}$ and
$A^{\dag\:2}$ are generators of SU(1,1), we multiply
$c_{1}A^{\dag\:2} + c_{2} A^{2}$ by $t$ and equate $\exp \left(
(c_{1}A^{\dag\:2} + c_{2} A^{2}) t \right)$ to the following
product of exponentials:
\begin{equation} \label{expressn}
\exp \left( f_{1}(t) A^{\dag\:2} \right) \exp \left( f_{2}(t)(A^{\dag}A +1/2)\right) \exp\left(f_{2}(t) A^{2}\right).
\end{equation}
Observe that in expression~(\ref{expressn}) that each exponent is
the product of an unknown function and a generator of $SU(1,1)$.
Furthermore, each of $SU(1,1)$'s generators appears exactly once.
Returning to the general case, finally, we calculate the unknown
functions of $t$, and so complete the process of disentangling the
single exponential, by differentiating both sides of the equation
equating the single exponential with the extra factor of $t$ in
its exponent to the product of exponentials with respect to $t$,
multiplying both sides from the right by the inverse of the single
exponential and, lastly, equating operator coefficients on both
sides. The paper \cite{truax88} contains a detailed example in
which the differential-equation approach is used.

%
%using d.e. approach to find identities
%

Using the BCH formula on the left-hand side of
Eq.~(\ref{identity1}) yields
\begin{eqnarray} \label{compose}
& & \hspace*{-1.7cm} \exp(c_{1} A^{2}) \exp(c_{2} A^{\dag} B^{\dag})
\\ \nonumber
& & \hspace*{-1.7cm} = \exp(c_{1}A^{2} + c_{2} A^{\dag} B^{\dag} + c_{1}c_{2} A B^{\dag}
+\frac{c_{1}c_{2}^{2}}{6} B^{\dag\:2}).
\end{eqnarray}
Noting that $\{A^{\dag} B^{\dag},A^{2}, A B^{\dag}, B^{\dag\:2}\}$
form a Lie algebra, we use the differential-equation approach on
the right-hand side of Eq.~(\ref{compose}) to know that there
exists a normally-ordered decomposition such that
\begin{eqnarray} \label{10_year_goal} \nonumber
\exp\bigl[ ( c_{1}A^{2} + c_{2} A^{\dag} B^{\dag} + c_{1}c_{2}
A B^{\dag} +\frac{c_{1}c_{2}^{2}}{6} B^{\dag\:2} )t \bigr]  & & \\
\nonumber & & \hspace*{-7.5cm} = \exp(f_{1}(t) A^{\dag} B^{\dag})
\exp(f_{2}(t) B^{\dag\:2}) \times \\
& & \hspace*{-7.0cm} \exp(f_{3}(t) B^{\dag} A) \exp(f_{4}(t) A^{2}) \\
\nonumber & & \hspace{-7.5cm} = U,
\end{eqnarray}
where $f_{1}$, $f_{2}$, $f_{3}$ and $f_{4}$ are functions we now
determine. Differentiating both sides of Eq.~(\ref{10_year_goal})
with respect to t and then multiplying from the right by $U^{-1}$
yields
\begin{eqnarray} \label{merz} \nonumber
& & c_{1}A^{2} + c_{2} A^{\dag} B^{\dag} + c_{1}c_{2} A B^{\dag}
+\frac{c_{1}c_{2}^{2}}{6} B^{\dag\:2} \\ \nonumber & & \;\; = \;\;
{\dot f}_{1} A^{\dag} B^{\dag} + {\dot f}_{2} \;e^{f_{1} A^{\dag}
B^{\dag}}\; B^{\dag\:2}\; e^{-f_{1}A^{\dag}B^{\dag}} \; \\
\nonumber & & + \; {\dot f}_{3} \; e^{f_{1}A^{\dag}B^{\dag}}\;
e^{f_{2} B^{\dag\:2}}\; B^{\dag} a \; e^{-f_{2} B^{\dag\:2}} \;
e^{-f_{1}A^{\dag}B^{\dag}} \;\; \\ \nonumber
& & + \;\; {\dot f}_{4} \;
e^{f_{1}A^{\dag}B^{\dag}}\; e^{f_{2} B^{\dag\:2}}\; e^{f_{3}
B^{\dag} A} \;B^{\dag\:2}  \; e^{-f_{3} B^{\dag} A} \\ \nonumber
& & \hspace*{-0.0cm}\times \;e^{-f_{2}
B^{\dag\:2}}\; e^{-f_{1} A^{\dag}B^{\dag}} \\
\end{eqnarray}
where ${\dot f_{i}}$ denotes $\partial f_{i}/ \partial t$.
Using the identity \cite{merzbacher61} (p. 162)
\begin{equation}
\exp({\cal A}) {\cal B} \exp(-{\cal A}) = {\cal B} + [{\cal A},{\cal B}]
+ 1/2! [{\cal A},[{\cal A},{\cal B}]] + \ldots,
\end{equation}
where ${\cal A}$ and ${\cal B}$ are arbitrary operators, on the
right-hand side of Eq.~(\ref{merz}) produces
\begin{eqnarray}
& & \hspace*{-1cm} c_{1}A^{2} + c_{2} A^{\dag} B^{\dag} + c_{1}c_{2} A B^{\dag}
+\frac{c_{1}c_{2}^{2}}{6} B^{\dag\:2}  \\ \nonumber
& & \hspace*{-2cm} = {\dot f}_{1} A^{\dag} B^{\dag} +
{\dot f}_{2} B^{\dag\:2}
+ {\dot f}_{3} (B^{\dag} A - f_{1} B^{\dag\:2}) \\ \nonumber
& & \hspace*{-2cm} + {\dot f}_{4} ( A^{2} - 2 f_{1}B^{\dag} A + f_{1}^{2}B^{\dag\:2} ).
\end{eqnarray}
Upon equating operator coefficients we arrive at four coupled differential equations
for the $f_{i}'s$. Solving these and setting t=1 yields
\begin{eqnarray}
& & \exp( c_{1}A^{2} + c_{2} A^{\dag} B^{\dag} + c_{1}c_{2} A B^{\dag}
+\frac{c_{1}c_{2}^{2}}{6} B^{\dag\:2}) \\ \nonumber
& & \exp(c_{2} A^{\dag}B^{\dag}) \;\exp(c_{1}c_{2}^{2} B^{\dag\:2})
\;\exp(2c_{1}c_{2}B^{\dag}A) \; \exp(c_{1} A^{2}).
\end{eqnarray}
Recalling that
\begin{eqnarray}
& & \exp( c_{1}A^{2} + c_{2} A^{\dag} B^{\dag} + c_{1}c_{2} A B^{\dag}
+\frac{c_{1}c_{2}^{2}}{6} B^{\dag\:2}) = \\ \nonumber
& & \exp(c_{1} A^{2}) \exp(c_{2} A^{\dag} B^{\dag})
\end{eqnarray}
we arrive at Eq.~(\ref{identity1}). The identity in
Eq.~(\ref{identity3}) can be obtained via a very similar
calculation to that which we have just performed.

%
%second identity SU(1,1)
%

To prove Eq.~(\ref{identity2}), we first note that $A^{2}$ and
$A^{\dag\:2}$ are generators of the Lie group SU(1,1). Related to
this group, it is known that \cite{schumaker}
\begin{eqnarray} \label{su_one_one} \nonumber
& & \hspace*{-0cm} \exp(c_{1} A^{2}) \exp(c_{3} (A^{\dag}A + 1/2))
\exp(c_{2} A^{\dag\:2})   \\ \nonumber
& & \hspace*{-0cm} = \exp(\beta_{1}(\{c_{i}\}) A^{\dag\:2}) \exp(\beta_{2}(\{c_{i}\}) (A^{\dag}A + 1/2))
\\
& & \hspace*{0.35cm} \exp(\beta_{3}(\{c_{i}\}) A^{\dag\:2}),
\end{eqnarray}
where $\beta_{1}$, $\beta_{2}$ and $\beta_{3}$ are, as yet,
unknown functions. Noting that these functions are only determined
by the commutation-relation structure of SU(1,1)'s generators, we
follow \cite{schumaker} and replace $A^{\dag\:2}$, $(A^{\dag}A +
1/2)$ and $A^{2}$ by two-dimensional matrices with identical
commutation relations. This leads us to making the following
transformations: $A^{\dag\:2}  \rightarrow 2 \sigma_{+}$, $A^{2}
\rightarrow - 2 \sigma_{-}$ and $A^{\dag} A + 1/2 \rightarrow
\sigma_{z}$, where
\begin{eqnarray}
\sigma_{+}=\left\{
\begin{array}{ll}
0 & 1 \\
0 & 0 \\
\end{array}
\right\},
\end{eqnarray}
\begin{eqnarray}
\sigma_{-}=\left\{
\begin{array}{ll}
0 & 0 \\
1 & 0 \\
\end{array}
\right\}
\end{eqnarray}
and
\begin{eqnarray}
\sigma_{z}=\left\{
\begin{array}{ll}
1 & 0 \\
0 & -1 \\
\end{array}
\right\} .
\end{eqnarray}
Upon doing this, and setting $c_{3}=0$, the left-hand side of
Eq.~(\ref{su_one_one}) becomes
\begin{eqnarray} \label{matrix_one}
\exp(- 2c_{1} \sigma_{-} )
\exp(2 c_{2} \sigma_{+})
& = & \left\{
\begin{array}{ll}
1 & 2c_{2} \\
 -2 c_{1} & 1-4c_{1}  c_{2} \\
\end{array}
\right\}
\end{eqnarray}
whilst the right-hand side transforms to
\begin{eqnarray} \label{matrix_two} \nonumber
& & \hspace*{-1cm} \exp(\beta_{1} 2\sigma_{+}) \exp(\beta_{2}\sigma_{z}) \exp(-\beta_{3} 2\sigma_{-}) \\
& & \hspace*{-1cm} =
\left\{
\begin{array}{ll}
P  -4 \beta_{1} \beta_{3} M & 2 \beta_{1} M \\
-2\beta_{3} M & M \\
\end{array}
\right\},
\end{eqnarray}
where $P=\cosh \beta_{2} + \sinh \beta_{2}$ and $M=\cosh \beta_{2}
- \sinh \beta_{2}$. Equating matrix elements on the right-hand
sides of Eqs~(\ref{matrix_one}) and (\ref{matrix_two}) leads to
\begin{eqnarray} \nonumber
\beta_{1}&=& c_{2}/(1 - 4c_{1} c_{2}) \\ \nonumber
\beta_{2}& =& \cosh^{-1} \bigl(1 + 2 c_{1}c_{2}/(1 - 4 c_{1} c_{2}) - 2 c_{1} c_{2}\bigr) \\ \nonumber
& {\rm and} & \\
\beta_{3}&=&c_{1}/(1 - 4 c_{1} c_{2})
\end{eqnarray}
and hence to Eq.~(\ref{identity2}).


\begin{thebibliography}{99}
\bibitem{num_errors} Numerical errors were induced in our values for $S_{2}$
as, practically, we had to truncate the sums in over $n_{1}$ and
$n_{2}$ in Eq.~(\ref{j_prob}) at finite values. These errors were
estimated by truncating the sums in at a number of increasing
$n_{1}$ and $n_{2}$ values and then seeing to what extent $S_{2}$
varied upon doing so.
\end{thebibliography}
\end{document}